\documentclass[%
 reprint,
 amsmath,amssymb,
prx,
floatfix,
superscriptaddress,
]{revtex4-2}

\usepackage{mathtools}
\usepackage{graphicx}
\usepackage{dcolumn}
\usepackage{bm}

\usepackage[caption=false]{subfig}

\usepackage{algorithm}
\usepackage[noend]{algpseudocode}
\algrenewcommand\algorithmicdo{}

\makeatletter
\renewcommand{\ALG@name}{Procedure}
\makeatother

\usepackage[linktocpage=true,
  colorlinks=true, 
  pdfborder={0 0 0},
  linkcolor=blue,
  citecolor=red,
  filecolor=yellow,
  urlcolor=blue,
  bookmarks,
  pdfauthor={},
]{hyperref}

\usepackage{ifthen}
\newcounter{is_qcircuit_used}
\newcounter{are_figs_merged}
\begin{document}

\preprint{APS/123-QED}

\title{
First-quantized eigensolver for ground and excited states of\\electrons under a uniform magnetic field
}

\author{Taichi Kosugi}
\email{kosugi.taichi@gmail.com}
\author{Hirofumi Nishi}
\affiliation{
Laboratory for Materials and Structures,
Institute of Innovative Research,
Tokyo Institute of Technology,
Yokohama 226-8503,
Japan
}

\affiliation{
Quemix Inc.,
Taiyo Life Nihombashi Building,
2-11-2,
Nihombashi Chuo-ku, 
Tokyo 103-0027,
Japan
}

\author{Yu-ichiro Matsushita}
\affiliation{
Laboratory for Materials and Structures,
Institute of Innovative Research,
Tokyo Institute of Technology,
Yokohama 226-8503,
Japan
}

\affiliation{
Quemix Inc.,
Taiyo Life Nihombashi Building,
2-11-2,
Nihombashi Chuo-ku, 
Tokyo 103-0027,
Japan
}

\affiliation{Quantum Material and Applications Research Center,
National Institutes for Quantum Science and Technology,
2-12-1, Ookayama, Meguro-ku, Tokyo 152-8552, Japan
}

\date{\today}

\begin{abstract}
First-quantized eigensolver (FQE) is a recently proposed quantum computation framework for obtaining the ground state of an interacting electronic system
based on probabilistic imaginary-time evolution.
Here, we propose a method for introducing a uniform magnetic field to the FQE calculation.
Our resource estimation demonstrates that the additional circuit responsible for the magnetic field can be implemented with a linear depth in terms of the number of qubits assigned to each electron.
Hence, introduction of the magnetic field has no impact on the leading order of the entire computational cost.
The proposed method is validated by numerical simulations of the ground and excited states employing filtration circuits for the energy eigenstates.
We also provide a generic construction of the derivative circuits together with measurement-based formulae.
As a special case of them,
we can obtain the electric-current density in an electronic system to gain insights into the microscopic origin of the magnetic response. 
\end{abstract}

\maketitle 

\section{Introduction}
\label{sec:introduction}

When an electronic system undergoes an external magnetic field,
interesting phenomena that are never seen without the field are known to emerge.
The most striking examples include the integer quantum Hall effect (IQHE) and the fractional QHE (FQHE) \cite{bib:5923, bib:5924, bib:5925, bib:5735, bib:5736} appearing in two-dimensional systems.
The IQHE originates from quantization of the in-plane motion of a single electron experiencing a perpendicular magnetic field leading to the discrete energy levels, called the Landau levels.
In contrast, the FQHE exhibits rational fractional quantum numbers stemming from the correlated motion of multiple electrons under a perpendicular magnetic field.

Besides the exotic phenomena mentioned above,
significant effects on practical applications of quantum systems are brought about by 
the coupling between an electronic system and an external magnetic field.
The most important one among them is chemical shifts in the energy splittings of nuclear spins in a molecule or a solid \cite{bib:5480}. 
Chemical shifts are mainly caused by the electron cloud partially shielding the external magnetic field near the nucleus.
Although chemical shifts depend on the electronic environment
and are smaller than the reference energy splittings in general,
they are not negligible for realizing precise quantum information processing based on nuclear magnetic resonance (NMR) techniques~\cite{bib:5916, bib:5917, bib:5918, bib:5919, bib:5920, bib:5921, bib:5922}.

Although the second-quantized formalism is widely adopted in the quantum computation community to determine the ground state of electronic systems,
the first-quantized formalism~\cite{bib:5373, bib:5372, bib:5328, bib:5824, bib:5737, bib:5658} (or equivalently the grid-based formalism)
has attracted recent attention as a promising alternative.
In particular, first-quantized eigensolver (FQE)~\cite{bib:5737},
which is a framework of quantum chemistry based on the probabilistic imaginary-time evolution (PITE),
provides nonvariational energy minimization with favorable scaling of computational resources compared to the second-quantized ones.
Exhaustive search for optimal molecular geometries is also possible within FQE~\cite{bib:geom_opt_in_PITE}.

Given the importance of understanding and exploiting the coupling between an electronic system and an external magnetic field as discussed above,
it is desirable to develop methods to incorporate external magnetic fields in the quantum computation of electronic systems.
Although such a method has already been proposed for the second-quantized formalism~\cite{bib:5926},
a first-quantized method remains lacking to our best knowledge.
In this study, we employ the operator-splitting technique used by Watanabe and Tsukada~\cite{bib:5624} for classical computers to efficiently incorporate a uniform magnetic field in the FQE calculation.
Our resource estimation demonstrates that the additional circuit responsible for the magnetic field can be implemented with a linear depth in terms of the number of qubits assigned to each electron,
hence there is no impact on the leading order of the entire computational cost.
In addition, we provide a generic construction of the derivative circuits, together with measurement-based formulae to extract the electric-current density that helps reveal the microscopic magnetic response of the target system.
We also propose filtration circuits for eliminating undesired energy eigenstates based on the idea of Chan et al.~\cite{bib:5658}.
Our new techniques are validated by performing FQE simulations for confined electronic systems under external magnetic fields.

\section{Methods}
\label{sec:methods}

\subsection{Brief review of PITE and first-quantized formalism}

\subsubsection{PITE circuits}

\begin{figure*}
\begin{center}
\includegraphics[width=14cm]{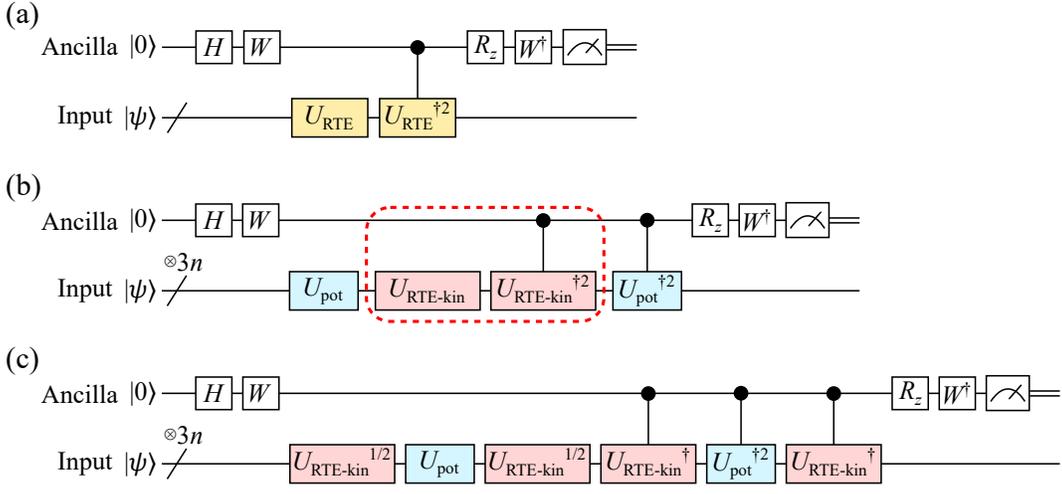}
\end{center}
\caption{
(a)
PITE circuit for a generic system governed by its Hamiltonian $\mathcal{H}.$
It contains
the RTE gate
$U_{\mathrm{RTE}} \equiv \exp(-i \mathcal{H} \Delta t)$
for an imaginary-time step $\Delta \tau$ and
$\Delta t \equiv s_1 \Delta \tau$.
$R_z \equiv R_z (-2 \theta_0)$ is the single-qubit $z$-rotation.
For the definition of $\theta_0$ and $W$,
see Ref.~\cite{bib:5737}
$H$ is the Hadamard gate.
(b) and (c) are
the circuits for a single particle when the $TV$ and $TVT$ splittings for $\mathcal{H}$ are adopted, respectively.
}
\label{fig:generic_approx_circuit}
\end{figure*}

The generic circuit for PITE \cite{bib:5737} is shown in
Fig.~\ref{fig:generic_approx_circuit}(a),
which uses a single ancilla to implement probabilistically the nonunitary imaginary-time evolution (ITE) operator
$\mathcal{M} = m_0 e^{-\mathcal{H} \Delta \tau}$
specified by the Hamiltonian $\mathcal{H}$ of a target system
and an imaginary-time step $\Delta \tau.$
$m_0$ is an adjustable real parameter.
PITE implements the desired nonunitary operation by resorting to a measurement.
(See also Refs.~\cite{Seki_Gutzwiller, 2022arXiv220209100M}.)
If the measurement outcome of the ancilla qubit is $| 0 \rangle$,
the evolved state for an arbitrary input state $| \psi \rangle$ within the first order of $\Delta \tau$ has been successfully obtained.
As seen in the circuit,
each PITE step is implemented by using the real-time evolution (RTE) gates
$U_{\mathrm{RTE}} (\Delta t) = e^{-i \mathcal{H} \Delta t}$
for the renormalized real-time step
$\Delta t \equiv s_1 \Delta \tau,$
where $s_1 \equiv m_0/\sqrt{1 - m_0^2}$ \cite{bib:5737}.
The PITE step has to be repeated successfully until the initial state becomes satisfactorily close to the ground state.
The step width does not need to be constant during the iterations.
For example, we can set it for the $k$th step to
\begin{gather}
    \Delta \tau_k
    =
        (1 - e^{-k/\kappa})
        (\Delta \tau_{\mathrm{max}} - \Delta \tau_{\mathrm{min}})
        +
        \Delta \tau_{\mathrm{min}}
        ,
    \label{def_variable_dtau}
\end{gather}
so that it changes gradually from $\Delta \tau_{\mathrm{min}}$
to $\Delta \tau_{\mathrm{max}}.$
$\kappa$ determines the rate of change.

\subsubsection{Encoding of many-particle wave functions}

We explain here the encoding of wave function in real space for multiple particles whose dynamics is governed by the Schr\"odinger equation \cite{bib:5373, bib:5372, bib:5328, bib:5824, bib:5737, bib:5658}.
We do not take spin degrees of freedom into account for simplicity.

We discretize a finite interval in each of $x, y,$ and $z$ directions into equidistant points to which the computational bases of $n$ qubits are assigned.
Specifically,
the $x$ coordinates of the $N \equiv 2^n$ points are given by
$x^{(k_x)} = k_x \Delta x \ (k_x = 0, \dots, N - 1)$,
where $\Delta x \equiv L/N$ is the grid step for an interval $[0, L]$ on the $x$ axis.
We assume that the simulation cell is a cube for simplicity.
The $y$ and $z$ coordinates are similarly given by
$y^{(k_y)} = k_y \Delta x$ and $z^{(k_z)} = k_z \Delta x,$ respectively.
We assign each computational basis
$
| \boldsymbol{k} \rangle_{3 n}
\equiv
| k_x \rangle_n \otimes | k_y \rangle_n \otimes | k_z \rangle_n$
of the $3 n$ qubits to each of the position eigenstates of each particle:
$
\hat{x} | \boldsymbol{k} \rangle_{3 n}
=
x^{(k_x)} | \boldsymbol{k} \rangle_{3 n}
,
$
where $\hat{x}$ is the position operator for $x$ direction.
We also write the eigenstate of $x$ position as $| x^{(k_x)} \rangle.$ 
The position operators and eigenstates for $y$ and $z$ directions are defined similarly.
Let us consider a system made up of $n_{\mathrm{par}}$ particles
and encode its wave function
$\Psi (\boldsymbol{r}_0, \dots, \boldsymbol{r}_{n_{\mathrm{par}} - 1})$
by using $3 n n_{\mathrm{par}}$ qubits.
We assume that the wave function is normalized to be unity when it is integrated over the simulation cell.
The coefficient of each computational basis thus represents the many-particle wave function as
\begin{gather}
    | \Psi \rangle
    =
        \Delta V^{n_{\mathrm{par}}/2}
        \sum_{
            \boldsymbol{k}_0,
            \dots, 
            \boldsymbol{k}_{n_{\mathrm{par}} - 1}
        }
        \Psi (
            \boldsymbol{r}^{(\boldsymbol{k}_0)},
            \dots,
            \boldsymbol{r}^{(\boldsymbol{k}_{n_{\mathrm{par}} - 1})}
        )
    \cdot
    \nonumber \\
    \cdot
        | \boldsymbol{k}_0
        \rangle_{3 n}
        \otimes \cdots \otimes
        | \boldsymbol{k}_{n_{\mathrm{par}} - 1} 
        \rangle_{3 n}
    ,
    \label{PITE_with_mag_fields:many_electron_state} 
\end{gather}
where we introduced the volume element $\Delta V \equiv (L/N)^3$
so that the normalization condition for the many-qubit state $\langle \Psi | \Psi \rangle = 1$ is satisfied.
In what follows,
we assume that the wave function has the periodicity of simulation cell for each direction of each position coordinate. 
We further assume that the initial many-particle state in an FQE calculation has already been symmetrized or antisymmetrized under exchange of any pair of particles according to the statistics \cite{bib:4825, bib:5389}.

As the canonical counterpart of the discretized positions,
we define the $N$ discrete momenta of a particle for each direction
$
p^{(\widetilde{s})}
\equiv
\widetilde{s}
\Delta p
\
(\widetilde{s} = -N/2, -N/2 + 1 \dots, N/2 - 1)
$
with the momentum step $\Delta p \equiv 2 \pi /L$ in reciprocal space.
The tilde symbol for an integer $j$ means $\widetilde{j} \equiv j - N/2$
in what follows.
We define the momentum eigenstate specified by three integers $s_x, s_y,$ and $s_z$
as the Fourier transform of the position eigenstates:
\begin{gather}
    | \boldsymbol{p}^{(\widetilde{ \boldsymbol{s} })} \rangle
    \equiv
        \frac{1}{N^{3/2}}
        \sum_{k_x = 0}^{N - 1}
        \sum_{k_y = 0}^{N - 1}
        \sum_{k_z = 0}^{N - 1}
        \exp 
        \left(
            i
            \boldsymbol{p}^{(\widetilde{ \boldsymbol{s} })}
            \cdot
            \boldsymbol{r}^{( \boldsymbol{k} )}
        \right)
        | \boldsymbol{k} \rangle_{3 n}
        .
    \label{QITE_as_a_part_of_RITE:def_mom_eigenstate}
\end{gather}
$| \boldsymbol{p}^{(\widetilde{ \boldsymbol{s} })} \rangle$
is also the eigenstate of
$\hat{T}_{0 \nu} \equiv \hat{p}_\nu^2/(2 m) \ (\nu = x, y, z)$
for each direction belonging to the discrete kinetic energy
$
E_{\mathrm{kin} s}
\equiv
\widetilde{s}^2 (\Delta p)^2/(2 m).
$
By defining the kinetic-phase gate $U_{\mathrm{kin}} (\Delta t)$
for $n$ qubits 
such that it acts on a computational basis $| j \rangle_n$ as
$
U_{\mathrm{kin}} (\Delta t)
| j \rangle_n
=
\exp ( -i E_{\mathrm{kin} j} \Delta t )
| j \rangle_n
,
$
the $n$-qubit RTE operator generated by
$\hat{T}_{0 \nu}$ can be implemented as 
\begin{align}
    \mathrm{CQFT}
    \cdot
    U_{\mathrm{kin}} (\Delta t)
    \cdot
    \mathrm{CQFT}^\dagger
    &=
        e^{-i \hat{T}_{0 \nu} \Delta t}
    \label{imag_evol_as_part_of_real_evol:CQFT_on_kinetic_rotation}
\end{align}
for each direction,
where CQFT is the centered quantum Fourier transform \cite{bib:5391, bib:5384, bib:5737}.
The consistent adoption of the range $[0, L]$ for spatial discretization and CQFT is just a matter of convention.
In fact, the combination of a range $[-L/2, L/2]$ and the ordinary QFT \cite{Nielsen_and_Chuang}
is possible in formulating the first-quantized approach \cite{bib:5658}.

\subsection{FQE for a particle under a uniform magnetic field}

\subsubsection{Hamiltonian}

In the present study,
we focus on the techniques for incorporating a uniform static magnetic field into the FQE framework.
We consider a single spinless quantum mechanical particle in three dimensional space.
It has a mass $m$ and a charge $q$ subject to an external potential $V$ and an external magnetic field $\boldsymbol{B}.$
The Hamiltonian $\mathcal{H}$ is the sum of the kinetic part
\begin{gather}
    \hat{T}
    =
        \frac{1}{2 m}
        \left(
            \hat{\boldsymbol{p}}
            -
            \frac{q}{c}
            \boldsymbol{A} (\hat{\boldsymbol{r}})
        \right)^2
    \label{def_kinetic_opr}
\end{gather}
and the potential part $\hat{V} \equiv V (\hat{\boldsymbol{r}}).$
$\hat{\boldsymbol{r}}$ is the position operator and
$\hat{\boldsymbol{p}}$ is the canonical-momentum operator.
The vector potential is defined such that its rotation gives the magnetic field:
$\boldsymbol{B} (\boldsymbol{r}) = \nabla \times \boldsymbol{A} (\boldsymbol{r}).$
For constructing the PITE circuit,
we have to implement the RTE operator
$
U_{\mathrm{RTE}} (\Delta t)
=
e^{-i (\hat{T} + \hat{V} ) \Delta t}
$
for the $3 n$ qubits
by employing the Suzuki--Trotter formula.
We examine in this study the following two splitting patterns,
that is,
the first-order splitting
\begin{gather}
    U_{\mathrm{RTE}} (\Delta t)
    =
        e^{-i \hat{T} \Delta t}
        e^{-i \hat{V} \Delta t}
        +
        \mathcal{O} (\Delta t^2)
        ,
\end{gather}
which we call the $TV$ splitting,
and the second-order splitting
\begin{gather}
    U_{\mathrm{RTE}} (\Delta t)
    =
        e^{-i \hat{T} \Delta t/2}
        e^{-i \hat{V} \Delta t}
        e^{-i \hat{T} \Delta t/2}
        +
        \mathcal{O} (\Delta t^3)
        ,
\end{gather}
which we call the $TVT$ splitting.
From the kinetic-evolution operator
$U_{\mathrm{RTE-kin}} (\Delta t) \equiv e^{-i \hat{T} \Delta t}$
and the potential-evolution operator
$U_{\mathrm{pot}} (\Delta t) \equiv e^{-i \hat{V} \Delta t},$
the PITE circuit for the $TV$ splitting is constructed
as shown in Fig.~\ref{fig:generic_approx_circuit}(b),
while that for the $TVT$ splitting is shown in Fig.~\ref{fig:generic_approx_circuit}(c).

\subsubsection{Kinetic evolution}

If there were no magnetic field,
the kinetic-evolution for the $x$, $y$, and $z$ directions in 
$U_{\mathrm{RTE-kin}} (\Delta t)$ could be decoupled.
This is not the case for the system we are considering.
We can assume without loss of generality
that the magnetic field is along the $z$ direction
and it is generated by the Landau gauge:
$\boldsymbol{A} (\boldsymbol{r}) = B x \boldsymbol{e}_y.$
With this choice of vector potential,
the kinetic evolution operator is decomposed as follows:
\begin{gather}
    U_{\mathrm{RTE-kin}} (\Delta t)
    \nonumber \\
    =
            \exp
            \left(
                -
                \frac{i \Delta t}{2 m}
                \left(
                    \hat{p}_y
                    -
                    \frac{q B}{c}
                    \hat{x}
                \right)^2
            \right)
        e^{-i \hat{T}_{0x} \Delta t}
        e^{-i \hat{T}_{0z} \Delta t}
        +
        \mathcal{O} (\Delta t^2)
        ,
    \label{PITE_with_mag_fields:splitting_kin_evol}
\end{gather}
where we used the first-order Suzuki--Trotter formula.
The term of the form $\hat{p}_y \hat{x} - \hat{x} \hat{p}_y$
involved in the exponent for $x$ and $y$ directions
is unfavorable for the implementation as a gate sequence.
Therefore we use the identity
$
e^{-i a (\hat{p}_y - b)^2}
=
e^{i b \hat{y}}
e^{-i a \hat{p}_y^2}
e^{-i b \hat{y}}
$
for arbitrary real $a$ and $b$ independent of $y$ to split
the evolution for $y$ direction exactly as
\begin{gather}
        \exp
        \left(
            -
            \frac{i \Delta t}{2 m}
            \left(
                \hat{p}_y
                -
                \frac{q B}{c}
                \hat{x}
            \right)^2
        \right)
        =
        U_{\mathrm{mag}}
        e^{-i \hat{T}_{0 y} \Delta t}
        U_{\mathrm{mag}}^\dagger
    ,
    \label{PITE_with_mag_fields:splitting_kin_evol_using_mag_phase}
\end{gather}
where
\begin{gather}
    U_{\mathrm{mag}}
    \equiv
        e^{i q B \hat{x} \hat{y}/c}
    \label{PITE_with_mag_fields:def_U_mag}
\end{gather}
is the magnetic-phase gate for the $2 n$ qubits,
independent of the time step.
The splitting in Eq.~(\ref{PITE_with_mag_fields:splitting_kin_evol_using_mag_phase})
is favorable since the position and momentum operators are decoupled.
This splitting was used by Watanabe and Tsukada \cite{bib:5624}
for simulations on a classical computer
to integrate the real-time-dependent Schr\"odinger equation for an electron under a magnetic field.
By substituting
Eq.~(\ref{PITE_with_mag_fields:splitting_kin_evol_using_mag_phase})
into
Eq.~(\ref{PITE_with_mag_fields:splitting_kin_evol}),
we find the easy-to-implement expression of the evolution operator:
\begin{gather}
    U_{\mathrm{RTE-kin}} (\Delta t)
    \nonumber \\
    =
        U_{\mathrm{mag}}
        e^{-i \hat{T}_{0 y} \Delta t}
        U_{\mathrm{mag}}^\dagger
        e^{-i \hat{T}_{0 x} \Delta t}
        e^{-i \hat{T}_{0z} \Delta t}
        +
        \mathcal{O} (\Delta t^2)
        ,
    \label{PITE_with_mag_fields:U_RTE_kin_xy_ST1}
\end{gather}
which is adopted in the present study.

\subsubsection{Gate sequences for PITE circuits}

\begin{figure*}
\begin{center}
\includegraphics[width=12cm]{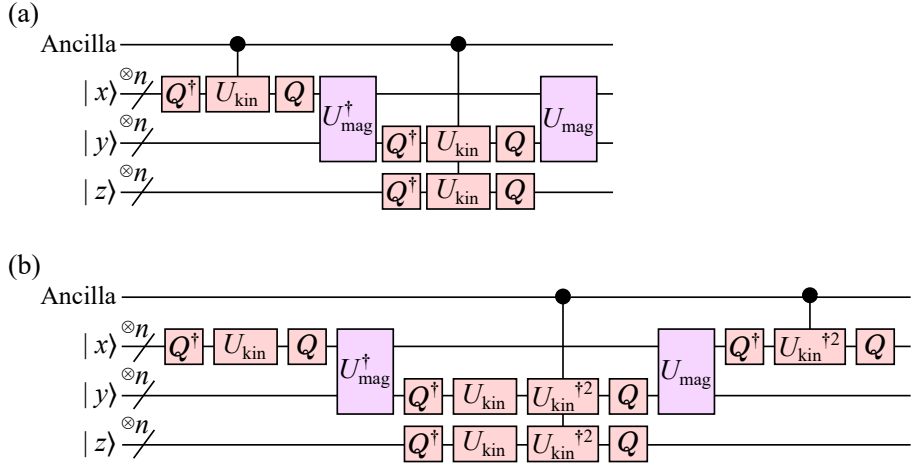}
\end{center}
\caption{
(a)
Gate sequence that implements
the controlled $U_{\mathrm{RTE-kin}} (\Delta t)$
in Eq.~(\ref{PITE_with_mag_fields:U_RTE_kin_xy_ST1}).
$Q$ represents the CQFT operator.
$U_{\mathrm{kin}} \equiv U_{\mathrm{kin}} (\Delta t)$
is used in the figure.
(b)
Implementation of the dashed region in
Fig.~\ref{fig:generic_approx_circuit}(b)
for the $TV$ splitting.
}
\label{fig:circuits_for_kinetic}
\end{figure*}

We can now find the gate sequences in the PITE circuit for the particle under the magnetic field
according to the splitting patterns.
As seen in Figs.~\ref{fig:generic_approx_circuit}(b) and (c),
$U_{\mathrm{RTE-kin}}$ in the PITE circuit can appear as a controlled gate.
It is easily confirmed that
$U_{\mathrm{RTE-kin}} (\Delta t)$
in Eq.~(\ref{PITE_with_mag_fields:U_RTE_kin_xy_ST1})
controlled by the ancilla
can be implemented as a gate sequence in 
Fig.~\ref{fig:circuits_for_kinetic}(a),
where only the kinetic-phase gates in 
Eq.~(\ref{imag_evol_as_part_of_real_evol:CQFT_on_kinetic_rotation})
are controlled.
For the $TV$ splitting of the total Hamiltonian,
the dashed region in Fig.~\ref{fig:generic_approx_circuit}(b)
is implemented as shown in 
Fig.~\ref{fig:circuits_for_kinetic}(b).
The PITE circuit for the $TVT$ splitting
is also constructed similarly (not shown).
It should be noted here that, 
whatever splitting pattern is adopted,
the error of the whole PITE calculation is $\mathcal{O} (\Delta t^2)$
since the original circuit in Fig.~\ref{fig:generic_approx_circuit}(a)
is correct within the first order of $\Delta t.$

The reason for the $TV$ splitting having been adopted by us instead of $VT$ splitting is now clear.
If the latter was adopted,
the cancellation of the parts of
$U_{\mathrm{RTE-kin}} (\Delta t)$ and 
$U_{\mathrm{RTE-kin}} (\Delta t)^{\dagger 2}$
would not have occurred and the circuits would be deeper than those in Figs.~\ref{fig:circuits_for_kinetic}(b).
It is similarly the case with the reason for the adoption of $TVT$ splitting instead of $VTV$ splitting.

\subsection{Implementation of phase gates}

\subsubsection{Kinetic-phase gate}

The kinetic-phase gate can be implemented by the technique
proposed by Benenti and Strini \cite{bib:5390}.
We describe the implementation briefly here in order for this paper to be self-contained.

We consider here a one-dimensional system for simplicity
where $n$ qubits are used per particle for encoding the wave function.
By defining $e_{\mathrm{kin}} \equiv \Delta p^2/(2 m)$ and 
the single-qubit phase gate
\begin{gather}
    Z_{\mathrm{kin}} (u)
    \equiv
    | 0 \rangle \langle 0 |
    +
    \exp
    \left( i u  e_{\mathrm{kin}} \Delta t \right)
    | 1 \rangle \langle 1 |
    \label{def_phase_kin_1q}
\end{gather}
specified by a real parameter $u$,
the kinetic-phase operator can be written as
\begin{gather}
    U_{\mathrm{kin}} (\Delta t)
    =
        e^{-i N^2 e_{\mathrm{kin}} \Delta t/4}
        \left(
            \prod_{\ell = 0}^{n - 1}
            \overbrace{
                Z_{\mathrm{kin}}
                (2^\ell (N - 2^{\ell}) )
            }^{\mathrm{On} \ \ell \mathrm{th \ qubit} }
        \right)
    \cdot
    \nonumber \\
    \cdot
            \prod_{\ell = 0}^{n - 1}
            \prod_{\ell' < \ell}
                \mathrm{C}_{\ell, \ell'}
                Z_{\mathrm{kin}}
                (-2^{\ell + \ell' + 1})
        ,
    \label{PITE_with_mag_fields:impl_of_U_kin}
\end{gather}
where $\mathrm{C}_{\ell, \ell'} Z_{\mathrm{kin}}$
represents the $Z_{\mathrm{kin}}$ gate on the $\ell'$th qubit controlled by the $\ell$th qubit.
For the derivation of Eq.~(\ref{PITE_with_mag_fields:impl_of_U_kin}),
see Appendix \ref{sec:derivation_of_kin_phase}.
We must not omit the overall phase factor in the expression
since the kinetic-phase gate in the PITE circuit can be controlled by the ancilla.
The gate sequence that implements $U_{\mathrm{kin}} (\Delta t)$ for $n = 4$
is shown in Fig.~\ref{fig:phase_gates_examples}(a) as an example.

\subsubsection{Magnetic-phase gate}

\begin{figure*}
\begin{center}
\includegraphics[width=13cm]{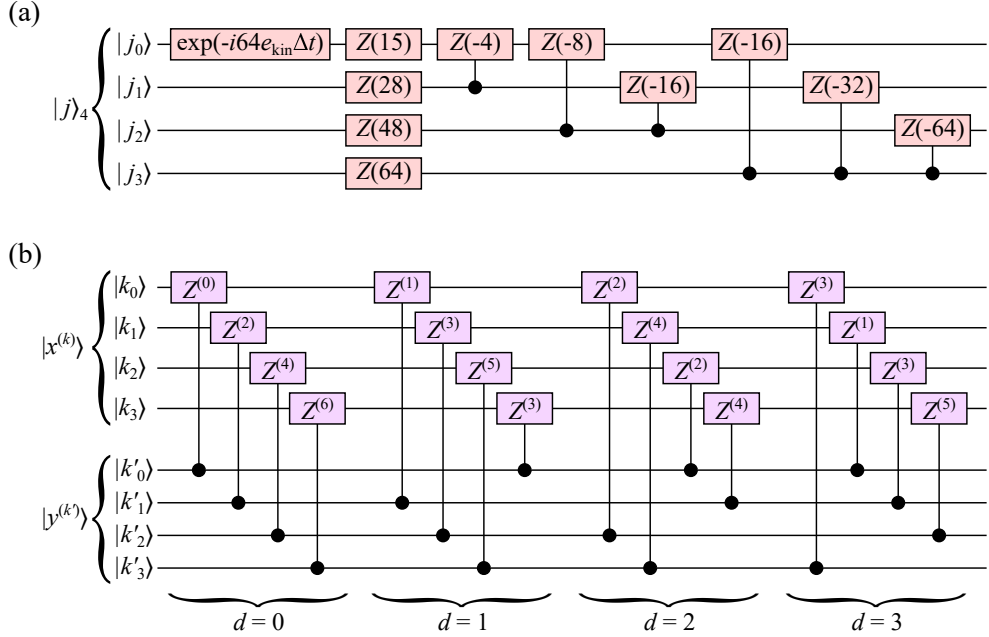}
\end{center}
\caption{
(a)
Gate sequence that implements the kinetic-phase gate
$U_{\mathrm{kin}} (\Delta t)$ for $n = 4.$
$Z (u)$ in this figure represents
$Z_{\mathrm{kin}} (u)$ defined in Eq.~(\ref{def_phase_kin_1q}).
(b)
Depth-$\mathcal{O} (n)$
gate sequence that implements the magnetic-phase gate
$U_{\mathrm{mag}}$ for $n = 4.$
$Z^{(\ell)}$ in this figure represents
$Z_{\mathrm{mag}}^{(\ell)}$ defined in Eq.~(\ref{def_phase_mag_1q}).
The values of $d$ correspond to those in
Eq.~(\ref{PITE_with_mag_fields:impl_of_U_mag}).
}
\label{fig:phase_gates_examples}
\end{figure*}

Here we describe the implementation of the magnetic-phase gate
defined in Eq.~(\ref{PITE_with_mag_fields:def_U_mag}).
From $\mu \equiv q B/c$,
we define the single-qubit phase gate
\begin{gather}
    Z_{\mathrm{mag}}^{(\ell)}
    \equiv
    | 0 \rangle \langle 0 |
    +
    \exp
    \left( i 2^\ell \mu \Delta x^2 \right)
    | 1 \rangle \langle 1 |
    \label{def_phase_mag_1q}
\end{gather}
specified by an integer $\ell.$
The magnetic-phase operator can then be written as
\begin{align}
    U_{\mathrm{mag}}
    =
        \prod_{d = 0}^{n - 1}
        \prod_{\ell = 0}^{n - 1}
        \mathrm{C}_{\ell, \ell'}
        Z_{\mathrm{mag}}^{(\ell + \ell')}
        |_{\ell' = \ell + d \ \mathrm{mod} \ n}
    ,
    \label{PITE_with_mag_fields:impl_of_U_mag}
\end{align}
where 
$\mathrm{C}_{\ell, \ell'} Z_{\mathrm{mag}}^{(\ell + \ell')}$
represents the $Z_{\mathrm{mag}}^{(\ell + \ell')}$ gate
on the $\ell'$th qubit for the $y$ position coordinate
controlled by the $\ell$th qubit for the $x$ position coordinate.
For the derivation of Eq.~(\ref{PITE_with_mag_fields:impl_of_U_mag}),
see Appendix \ref{sec:derivation_of_mag_phase}.
Equation~(\ref{PITE_with_mag_fields:impl_of_U_mag})
enables one to implement $U_{\mathrm{mag}}$
as a depth-$\mathcal{O}(n)$ circuit
since, for a fixed $d$, the controlled single-qubit phase gates can be parallelized.
The gate sequence that implements $U_{\mathrm{mag}}$ for $n = 4$
is shown in Fig.~\ref{fig:phase_gates_examples}(b) as an example.

We can also adopt the shifted Landau gauge $\boldsymbol{A} (\boldsymbol{r}) = (x - x_{\mathrm{g}}) B \boldsymbol{e}_y$ with an arbitrary $x_{\mathrm{g}},$ leading to the same magnetic field as the unshifted one.
The magnetic-phase gate in such a case is
$U_{\mathrm{mag}} = e^{i q B (\hat{x} - x_{\mathrm{g}}) \hat{y}/c}$
instead of Eq.~(\ref{PITE_with_mag_fields:def_U_mag}),
which requires us to append the circuit implementation of
$e^{-i q B x_{\mathrm{g}} \hat{y}/c}$ to that in
Eq.~(\ref{PITE_with_mag_fields:impl_of_U_mag}).
Since the depth of the additional circuit is $\mathcal{O} (1),$
the scaling of the entire magnetic-phase circuit is the same as for the unshifted gauge.

\subsection{Computational cost}

In this subsection,
we compare the splitting patterns for the RTE operator
to estimate their computational cost for PITE calculations.

For a single particle in three-dimensional space,
Table~\ref{table:PITE_with_mag_fields:opr_numbers} shows
the numbers of calls of major subroutines in a single PITE step
[see Figs.~\ref{fig:generic_approx_circuit}(b) and (c)]
for the two splitting patterns with or without a magnetic field.
The call number of $U_{\mathrm{pot}}$ is independent of the presence of a magnetic field and the order ($TV$ or $TVT$) of splitting.
The call number of $U_{\mathrm{mag}}$ is smaller than those of
QFT and $U_{\mathrm{kin}}.$
Furthermore, as stated above, $U_{\mathrm{mag}}$ admits the depth-$\mathcal{O}(n)$ implementation,
whereas QFT and $U_{\mathrm{kin}}$ require $\mathcal{O}(n^2)$ depths.
The operations of $U_{\mathrm{mag}}$ thus does not contribute to
the scaling of total depth with respect to $n.$

\begin{figure*}
\begin{center}
\includegraphics[width=17cm]{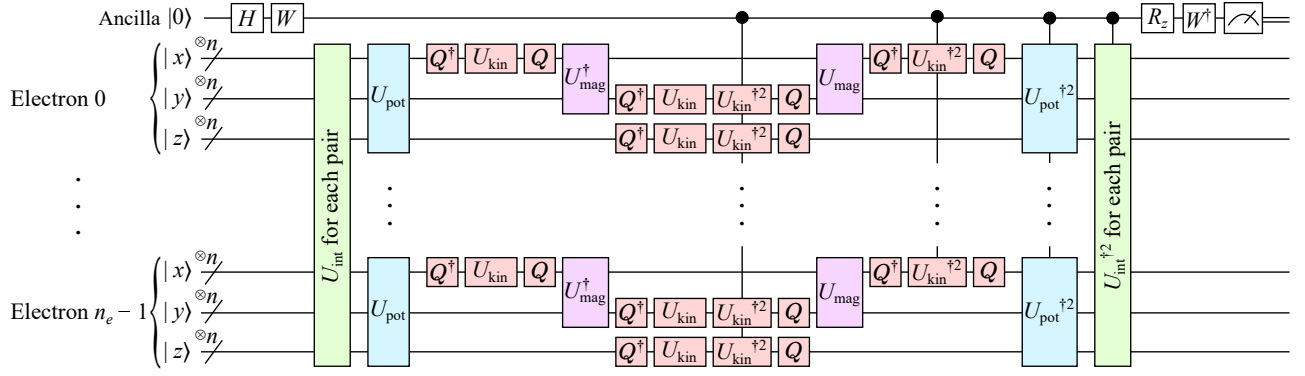}
\end{center}
\caption{
Circuit for a single PITE step used in an FQE calculation for $n_e$ interacting electrons under a static magnetic field.
This circuit involves the $TV$ splitting circuit in
Fig.~\ref{fig:circuits_for_kinetic}(b) for each electron.
}
\label{fig:circuit_for_electrons}
\end{figure*}

In a practical system where $n_e$ electrons interact with each other,
particle-wise parallelization of $U_{\mathrm{mag}}$ and QFT operations is possible in a single PITE step,
whereas that of $U_{\mathrm{kin}}, U_{\mathrm{pot}},$
and the phase gate $U_{\mathrm{int}}$ for the interactions \cite{bib:5737}
is not possible due to the ancillary qubit.
Therefore $U_{\mathrm{mag}}$ does not contribute to
the scaling of total depth with respect to $n_e.$
The circuit for a single PITE step used in such an FQE calculation for a molecule is depicted in
Fig.~\ref{fig:circuit_for_electrons}.
This circuit is a many-electron extension of that in
Fig.~\ref{fig:generic_approx_circuit}(b)
and involves the $TV$ splitting circuit in
Fig.~\ref{fig:circuits_for_kinetic}(b) for each electron.
This is also a nonzero-magnetic-field version of the circuit in Fig.~2(b) in Ref.~\cite{bib:5737}.
The PITE circuit for a molecule using the $TVT$ splitting
can also be constructed similarly (not shown).
Also,
the introduction of a magnetic field
does not increase the call numbers of
$U_{\mathrm{kin}}, U_{\mathrm{pot}},$ and $U_{\mathrm{int}}.$
In short, the introduction of a magnetic field to an interacting electronic system does not lead to a large increase in computational cost.
This is also the case where
an external magnetic field is introduced for the geometry optimization within FQE \cite{bib:geom_opt_in_PITE}.

Here we mention the comparison with second-quantized approaches.
For incorporating an external static magnetic field into a second-quantized Hamiltonian in quantum chemistry on a classical computer,
one often employs the one-electron atomic orbitals that take the presence of the vector potential into account \cite{bib:4063, bib:3349}.
The one- and two-electron integrals between them are calculated for the parameters in the Hamiltonian.
Once it is constructed,
the classical computation for finding the ground state for the fixed molecular geometry proceeds similarly to no-magnetic-field cases.
The Hamiltonian can also be adopted for a second-quantized approach on a quantum computer.
The advantages and disadvantages of the first- and second-quantized approaches over each other in no-magnetic-field cases \cite{bib:5737} are inherited to nonzero-magnetic-field cases in principle.
For example,
a second-quantized approach requires much less number of qubits for encoding electronic orbitals in general, while a first-quantized approach requires qubits for sufficiently many grid points for each electron. 
On the other hand,
a second-quantized approach requires $\mathcal{O} (n_e^5)$ classical computation for the two-electron integrals between the molecular orbitals before the quantum computation starts.
In addition, the interaction terms in the Hamiltonian demand $\mathcal{O} (n_e^4)$ quantum computational time.
Our first-quantized approach is free from both of them.

\begin{table}
\centering
\caption{Numbers of calls of major subroutines in a single PITE step for a single particle. Those in parentheses represent the call numbers of controlled subroutines.}
\begin{tabular}{ccccl} 
\hline
Splitting pattern    & QFT                  & $U_{\mathrm{kin}}$   & $U_{\mathrm{mag}}$   & $U_{\mathrm{pot}}$         \\ 
\hline
$B=0$                & \multicolumn{1}{l}{} & \multicolumn{1}{l}{} & \multicolumn{1}{l}{} &                            \\
$TV$                 & 6                    & 6 (3)                & 0                    & \multicolumn{1}{c}{2 (1)}  \\
$TVT$                & 18                   & 12 (6)               & 0                    & \multicolumn{1}{c}{2 (1)}  \\
\multicolumn{1}{l}{} & \multicolumn{1}{l}{} & \multicolumn{1}{l}{} & \multicolumn{1}{l}{} &                            \\
$B \ne 0$            & \multicolumn{1}{l}{} & \multicolumn{1}{l}{} & \multicolumn{1}{l}{} &                            \\
$TV$                 & 8                    & 6 (3)                & 2                    & \multicolumn{1}{c}{2 (1)}  \\
$TVT$                & 20                   & 12 (6)               & 6                    & \multicolumn{1}{c}{2 (1)}  \\
\hline
\end{tabular}
\label{table:PITE_with_mag_fields:opr_numbers}
\end{table}

\subsection{Obtaining electric-current density via measurements}

\subsubsection{One-electron density matrix and electric-current density}

For the $n_e$-electron state $| \Psi \rangle$ represented by
Eq.~(\ref{PITE_with_mag_fields:many_electron_state}),
the reduced density operator \cite{Nielsen_and_Chuang}
for a single electron is calculated by
tracing out the degrees of freedom of $n_e - 1$
electrons from the total density operator:
\begin{align}
    \rho_{\mathrm{one}}
    &=
        \mathrm{Tr}_{n_e - 1}
        | \Psi \rangle \langle \Psi |
    \nonumber \\
    &=
        \frac{\Delta V}{n_e}
        \sum_{\boldsymbol{k}, \boldsymbol{k}'}
            \gamma
            (
                \boldsymbol{r}^{(\boldsymbol{k})},
                \boldsymbol{r}^{(\boldsymbol{k}')}
            )
            | \boldsymbol{k} \rangle
            \langle \boldsymbol{k}' |_{3 n}
    ,
    \label{def_reduced_dens_opr}
\end{align}
where
\begin{gather}
    \gamma (\boldsymbol{r}, \boldsymbol{r}')
    \equiv
        n_e
        \int
        d^3 \boldsymbol{r}_1
        \cdots
        d^3 \boldsymbol{r}_{n_e - 1}
        \Psi (
            \boldsymbol{r},
            \boldsymbol{r}_1,
            \dots,
            \boldsymbol{r}_{n_e - 1}
        )
    \cdot
    \nonumber \\
    \cdot
        \Psi (
            \boldsymbol{r}',
            \boldsymbol{r}_1,
            \dots,
            \boldsymbol{r}_{n_e - 1}
        )^*
    \nonumber \\
    =
        n_e
        \Delta V^{n_e - 1}
        \sum_{
            \boldsymbol{k}_1,
            \dots, 
            \boldsymbol{k}_{n_e - 1}
        }
        \Psi (
            \boldsymbol{r},
            \boldsymbol{r}^{(\boldsymbol{k}_1)},
            \dots,
            \boldsymbol{r}^{(\boldsymbol{k}_{n_e - 1})}
        )
    \cdot
    \nonumber \\
    \cdot
        \Psi (
            \boldsymbol{r}',
            \boldsymbol{r}^{(\boldsymbol{k}_1)},
            \dots,
            \boldsymbol{r}^{(\boldsymbol{k}_{n_e - 1})}
        )^*
    \label{def_reduced_dens_mat_1e}
\end{gather}
is the reduced density matrix \cite{stefanucci2013nonequilibrium}
for a single electron,
or equivalently the one-electron density matrix (1DM),
in position representation.
Its normalization condition we have adopted is the conventional one in condensed-matter physics.
Specifically,
the diagonal components give the electron density $\rho (\boldsymbol{r})$ to satisfy
\begin{gather}
    \int
    d \boldsymbol{r}
        \gamma (\boldsymbol{r}, \boldsymbol{r})
    =
        \Delta V
        \sum_{\boldsymbol{k}}
            \rho (\boldsymbol{r}^{(\boldsymbol{k})})
    =
        n_e
    ,
\end{gather}
which leads to the normalization condition
$\mathrm{Tr} \rho_{\mathrm{one}} = 1$
for Eq.~(\ref{def_reduced_dens_opr}),
the conventional one in quantum information processing.

It is known that
the paramagnetic electric-current density as a function of position can be calculated from the 1DM as \cite{stefanucci2013nonequilibrium}
\begin{gather}
    \boldsymbol{j}_{\mathrm{para}}
    (\boldsymbol{r})
    =
        \frac{(-e)}{m}
        \mathrm{Im}
            \frac{\partial
                \gamma (\boldsymbol{r}, \boldsymbol{r}')}{
                \partial \boldsymbol{r}
            }
            \Bigg|_{\boldsymbol{r}' = \boldsymbol{r}}
    \label{current_dens_paramag_using_RDM1}
\end{gather}
and the diamagnetic electric-current density can be calculated as
\begin{gather}
    \boldsymbol{j}_{\mathrm{dia}} (\boldsymbol{r})
    =
        -
        \frac{(-e)^2}{m c}
        \rho (\boldsymbol{r})
        \boldsymbol{A} (\boldsymbol{r})
    .
    \label{current_dens_diamag_using_RDM1}
\end{gather}
The factor $-e < 0$ in 
Eqs.~(\ref{current_dens_paramag_using_RDM1}) and
(\ref{current_dens_diamag_using_RDM1})
comes from the negative electron charge.
It is noted that the paramagnetic and diamagnetic current densities are solely gauge dependent,
whereas the total current density
$
\boldsymbol{j} (\boldsymbol{r})
=
    \boldsymbol{j}_{\mathrm{para}} (\boldsymbol{r})
    +
    \boldsymbol{j}_{\mathrm{dia}} (\boldsymbol{r})
$
is gauge invariant to admit physically meaningful interpretations \cite{stefanucci2013nonequilibrium}.

\subsubsection{Circuits and measurement-based formula}

The electric-current density provides insights for understanding the microscopic origin of the response of a target system to an external magnetic field such as the shielding due to the electron clouds \cite{bib:4247, bib:5720}.
We describe below how the electric-current density can be obtained from measurements on qubits that encode the wave function in real space.

The electron density can be obtained simply by
measuring $3 n$ qubits in $| \Psi \rangle$
since the electrons are indistinguishable.
Specifically,
the electron density at $\boldsymbol{r}^{(\boldsymbol{k})}$
is related to 
the probability for observing
$| \boldsymbol{k} \rangle_{3 n}$ for an arbitrary electron among $n_e$ as
\begin{gather}
    \rho (\boldsymbol{r}^{(\boldsymbol{k})})
    =
        \frac{n_e}{\Delta V}
        \mathbb{P}_{\boldsymbol{k}}
        .
    \label{electron_dens_from_meas}
\end{gather}
The electron density constructed via repeated measurements
and 
Eq.~(\ref{current_dens_diamag_using_RDM1})
allow us to get the diamagnetic current density.

\begin{figure}
\begin{center}
\includegraphics[width=8.5cm]{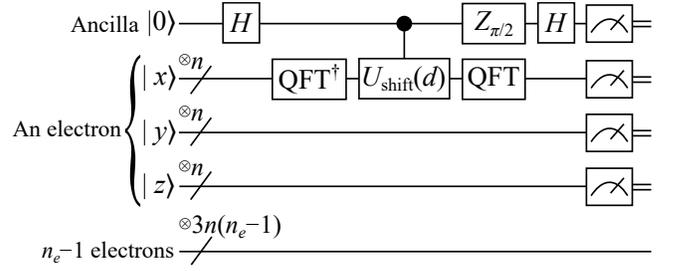}
\end{center}
\caption{
$(3 n + 1)$-qubit circuit $\mathcal{C}_{\mathrm{para}, x} (d)$ for
obtaining the $x$ component of paramagnetic electric-current density $\boldsymbol{j}_{\mathrm{para}} (\boldsymbol{r})$.
$
Z_{\pi/2}
= 
| 0 \rangle \langle 0 | + e^{i \pi/2} | 1 \rangle \langle 1 |
$
is a single-qubit phase gate.
The measurements are performed on the ancillary qubit and the $3 n$ qubits responsible for an arbitrary electron among $n_e.$
}
\label{fig:para_current_density_x}
\end{figure}

In contrast to the diamagnetic current density,
obtaining the paramagnetic current density requires some trick due to the presence of the position derivative of 1DM.
[See Eq.~(\ref{current_dens_paramag_using_RDM1})]
For obtaining the $x$ component of paramagnetic current density,
we introduce the circuit $\mathcal{C}_{\mathrm{para}, x} (d),$
as shown in Fig.~\ref{fig:para_current_density_x}.
This circuit comes from the generic method
for obtaining the derivative distributions
of a complex function described in
Appendix~\ref{sec:fdfc_distr_from_measurements}.
This circuit acts on the ancilla and the $3 n$ qubits for an arbitrary electron among the $n_e$ electrons. 
$U_{\mathrm{shift}} (d)$ is a phase gate
specified by an integer $d$
which acts on the computational basis as
$
| j \rangle_n
\xmapsto{}
\exp(-i 2 \pi d j/N )
| j \rangle_n
\
(j = 0, \dots, N - 1)
.
$
The circuit ends up with a measurement on the ancilla and
the $3 n$ qubits responsible for the single electron
with ignoring the other qubits.

We use the circuits 
$\mathcal{C}_{\mathrm{para}, x} (d)$ and
$\mathcal{C}_{\mathrm{para}, x} (-d)$ for obtaining the desired quantity.
Specifically, we have the measurement-based formula
\begin{gather}
    j_{\mathrm{para}, x} (\boldsymbol{r}^{(\boldsymbol{k})})
    =
        \frac{(-e)}{m}
        \frac{1}{2 d \Delta x}
        \Bigg(
            \frac{2 n_e}{\Delta V}
            \left(
                -
                \mathbb{P}_{\boldsymbol{k} 0} (-d)
                +
                \mathbb{P}_{\boldsymbol{k} 0} (d)
            \right)
    \nonumber \\
            +
            \frac{
                \rho
                (
                    \boldsymbol{r}^{(\boldsymbol{k})}
                    + d \Delta x \boldsymbol{e}_x
                )
                -
                \rho
                (
                    \boldsymbol{r}^{(\boldsymbol{k})}
                    - d \Delta x \boldsymbol{e}_x
                )
            }{2}
        \Bigg)
        +
        \mathcal{O} (\Delta x^2)
        ,
    \label{current_dens_paramag_x_from_meas}
\end{gather}
where $\mathbb{P}_{\boldsymbol{k} 0} (\pm d)$ is the probability for
observing $| \boldsymbol{k} \rangle_{3 n}$ for the electron
and $| 0 \rangle$ for the ancilla
when a measurement is performed for the $\mathcal{C}_{\mathrm{para}, x} (\pm d)$ circuit.
Equation~(\ref{current_dens_paramag_x_from_meas})
enables one to calculate the $x$ component of paramagnetic current density via repeated measurements
within the first order of $\Delta x.$
For the derivation of this equation,
see Appendix~\ref{sec:current_dens_paramag_x_from_meas}.
The circuits
$\mathcal{C}_{\mathrm{para}, y} (d)$ and
$\mathcal{C}_{\mathrm{para}, z} (d)$ for obtaining
the $y$ and $z$ components, respectively,
can also be constructed similarly.

\subsection{Filtration circuits for excited states}
\label{sec:filtration}

Since any kind of ITE approach lets the lowest-energy eigenstate in an initial state
survive imaginary-time lapses
while it forces the other eigenstates to diminish,
it is desirable to develop techniques
that remove a specific energy eigenstate from the initial state \cite{bib:5658}.
To this end, we describe below two circuits for probabilistic filtration,
which require only the estimated energy eigenvalues and does not require the eigenstates.
There can often be a case in which
we have estimated lowest energy eigenvalues
via some reliable approaches such as those related to Krylov subspace \cite{bib:5665, bib:4959, bib:5359, bib:5661},
without having obtained the eigenstates as qubits.

We should point out here that,
while the schemes below are for killing a single undesirable state via a single measurement,
Meister and Benjamin \cite{bib:5915} have already proposed elaborate schemes for killing multiple undesirable states via consecutive measurements based on the knowledge of energy levels within a target range.

\subsubsection{First-order filtration}

For an input $n$-qubit state $| \psi \rangle$ whose Hamiltonian is $\mathcal{H}$,
we construct the first-order probabilistic filtration circuit 
$\mathcal{C}_{\mathrm{filt1}}^{(\lambda)}$
for a target elimination energy $\lambda,$
as shown in Fig.~\ref{fig:filtration_circuits}(a).
This circuit is essentially the same as the one proposed by Chan et al.~\cite{bib:5658}
The real-time step $\Delta t_{\mathrm{f}}$ is an adjustable parameter.
The state of $(n + 1)$-qubit system immediately before the measurement on the ancilla is
\begin{gather}
        \frac{
        e^{-i \lambda \Delta t_{\mathrm{f}} /2}
        -
        e^{i (\lambda/2 - \mathcal{H}) \Delta t_{\mathrm{f}} }}{2}
        | \psi \rangle
        \otimes
        | 0 \rangle
    \nonumber \\
        +
        \frac{
        e^{-i \lambda \Delta t_{\mathrm{f}} /2}
        +
        e^{i (\lambda/2 - \mathcal{H}) \Delta t_{\mathrm{f}} }}{2}
        | \psi \rangle
        \otimes
        | 1 \rangle
        .
    \label{eigenstate_canceller:action_of_filt}
\end{gather}
Let us consider a practical situation
in which the observer wants to remove the ground state
$| \phi_0 \rangle$ from the input state
with knowing an estimated ground state energy $\widetilde{E}_0$ to obtain the first excited state $| \phi_1 \rangle$ in subsequent procedures.
The estimated energy can have an error $\delta E_0$ from the true energy eigenvalue $E_0$ as
$\widetilde{E}_0 = E_0 + \delta E_0$
and the observer uses $\mathcal{C}_{\mathrm{filt1}}^{(\widetilde{E}_0)}.$
Such errors can stem from various origins, e.g.,
a finite number of measurements to estimate the energy eigenvalue or a dephasing process the qubits undergo or
classical noises disturbing the circuit parameters.
The explanations of filtration techniques below,
however,
are not affected by what the origins are.
It is clear from
Eq.~(\ref{eigenstate_canceller:action_of_filt})
that, if the error is zero and the measurement outcome is $| 0 \rangle$,
the resultant $n$-qubit state,
which we call the success state in what follows,
does not contain $| \phi_0 \rangle.$

If the observer has an estimated energy $\widetilde{E}_1$
of the first excited state
that can have an error as well as the ground state, 
it may be appropriate to set the time step $\Delta t_{\mathrm{f}}$ to
\begin{gather}
    \Delta t_1
    \equiv
        \frac{\pi}{\widetilde{E}_1 - \widetilde{E}_0}
        .
    \label{eigenstate_canceller:optimal_dt}
\end{gather}
For this time step,
the weight of $| \phi_1 \rangle$ in the success state
acquires a factor $\sin^2 (\pi/2) = 1$ if the error $\delta E_1$ is zero.
Specifically,
by expanding the input state in terms of the energy eigenstates as
$| \psi \rangle = \sum_k c_k | \phi_k \rangle$,
the unnormalized success state is written as
(see Appendix \ref{sec:derivation_for_filtration})
\begin{gather}
    \sum_k
        i
        c_k
        \exp
        \left(
            -i
            \frac{\pi E_k}
            {2 (\widetilde{E}_1 - \widetilde{E}_0)}
        \right)
        \sin
        \frac{\pi (E_k - \widetilde{E}_0)}
        {2 (\widetilde{E}_1 - \widetilde{E}_0)}
        | \phi_k \rangle
        ,
    \label{eigenstate_canceller:success_state_for_filt1}
\end{gather}
where the amplitude of ground state is $\mathcal{O} (\delta E_0).$
The relative weight of the ground state compared to the first excited state in the success state is,
from
Eq.~(\ref{eigenstate_canceller:success_state_for_filt1}),
\begin{align}
    r_{\mathrm{filt 1}}
    \equiv
        \left| \frac{c_0}{c_1} \right|^2
        \frac{
            \sin^2
            \frac{\pi \overline{\delta} E_0}{2 (1 + \overline{\delta} E_1 - \overline{\delta} E_0)}
        }{
            \cos^2
            \frac{\pi \overline{\delta} E_1 }{2 (1 + \overline{\delta} E_1 - \overline{\delta} E_0)}
        }
    ,
\end{align}
where
$\overline{\delta} E_k \equiv \delta E_k/(E_1 - E_0) \ (k = 0, 1)$
is the relative error with respect to the energy gap.

After the first-order filtration described above,
the PITE steps begin
from the $n$-qubit state having a possibly small weight of the ground state
to eliminate the second and higher excited states.
In order for the resultant state after 
$n_{\mathrm{steps}}$ PITE steps to be close to the first excited state,
the relative weight
$\exp (n_{\mathrm{steps}} (E_1 - E_0) \Delta \tau) r_{\mathrm{filt 1}}$
of the ground state has to be still small, say $\varepsilon.$
The condition for this requirement is 
\begin{align}
    n_{\mathrm{steps}}
    \Delta \tau
    <
        \frac{1}{E_1 - E_0}
        \Bigg(
            \ln \varepsilon
            +
            2
            \ln
            \left| \frac{c_1}{c_0} \right|
    \nonumber \\
            +
            2
            \ln
            \left|
            \frac{
                \cos
                \frac{\pi \overline{\delta} E_1 }{2 (1 + \overline{\delta} E_1 - \overline{\delta} E_0)}
            }{
                \sin
                \frac{\pi \overline{\delta} E_0}{2 (1 + \overline{\delta} E_1 - \overline{\delta} E_0)}
            }
            \right|
        \Bigg)
        \equiv
        \tau_{\mathrm{ub1}}
        ,
    \label{eigenstate_canceller:cond_for_small_psi0_after_ITE_steps}
\end{align}
which imposes the upper bound $\tau_{\mathrm{ub1}}$
on the allowed total imaginary-time lapse.
In particular,
when the errors are smaller than the energy gap
($\overline{\delta} E_0 \ll 1$ and
$\overline{\delta} E_1 \ll 1$),
the upper bound is approximated as
\begin{align}
    \tau_{\mathrm{ub 1}}
    \approx
        \frac{1}{E_1 - E_0}
        \left(
            \ln \frac{4 \varepsilon}{\pi^2}
            +
            2
            \ln
            \left| \frac{c_1}{c_0} \right|
            +
            2
            \ln
            \frac{1}{| \overline{\delta} E_0 |}
        \right)
        .
    \label{filtration_ub1_approx}
\end{align}
This expression tells us that, for a fixed tolerance $\varepsilon$,
a better initial state ($| c_1 | \gg | c_0 |$)
and a better estimation for the ground state energy ($| \delta E_0 | \ll | E_1 - E_0 |$)
lead to a larger upper bound.
When the lapse exceeds the upper bound,
we will have to apply the filtration again to get the target excited state.

\subsubsection{Second-order filtration}

\begin{figure}
\begin{center}
\includegraphics[width=8.5cm]{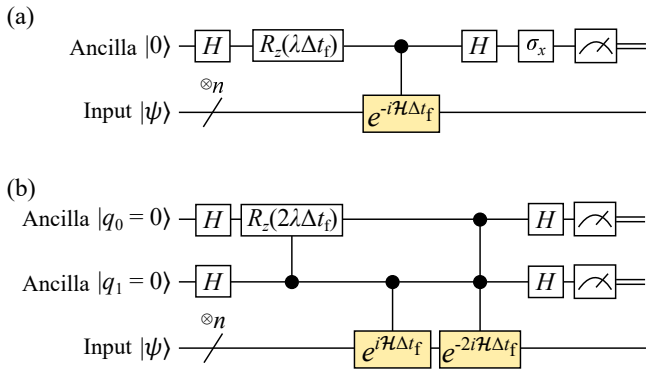}
\end{center}
\caption{
Circuits
(a)
$\mathcal{C}_{\mathrm{filt1}}^{(\lambda)}$
for first-order filtration and
(b)
$\mathcal{C}_{\mathrm{filt2}}^{(\lambda)}$
for second-order filtration
specified by a target energy $\lambda$.
$\Delta t_{\mathrm{f}}$ is an adjustable real-time step.
If $\lambda$ is equal to one of the energy eigenvalues
and the measurement outcome
$| 0 \rangle$ ($| 1 \rangle \otimes | 0 \rangle$)
for
$\mathcal{C}_{\mathrm{filt1}}^{(\lambda)}$
($\mathcal{C}_{\mathrm{filt2}}^{(\lambda)}$)
is obtained,
the energy eigenstate belonging to $\lambda$ has been filtered out
from the input state.
}
\label{fig:filtration_circuits}
\end{figure}

By introducing two ancillae $| q_1 \rangle \otimes | q_0 \rangle$,
we can construct the second-order probabilistic filtration circuit
$\mathcal{C}_{\mathrm{filt2}}^{(\lambda)}$
for the input $| \psi \rangle,$
as shown in Fig.~\ref{fig:filtration_circuits}(b).
The state of $(n + 2)$-qubit system immediately before the measurement on the ancillae is
\begin{gather}
        \cos^2
        \frac{ (\mathcal{H} - \lambda) \Delta t_{\mathrm{f}}}{2}
        | \psi \rangle \otimes | 0 \rangle \otimes | 0 \rangle
    \nonumber \\
        +
        \sin^2
        \frac{ (\mathcal{H} - \lambda) \Delta t_{\mathrm{f}}}{2}
        | \psi \rangle \otimes | 1 \rangle \otimes | 0 \rangle
    \nonumber \\
    +
        \frac{i}{2}
        \sin ((\mathcal{H} - \lambda) \Delta t_{\mathrm{f}})
        | \psi \rangle
        \otimes
        \left( | 0 \rangle - | 1 \rangle \right)
        \otimes
        | 1 \rangle
        ,
    \label{eigenstate_canceller:action_of_filt_2}
\end{gather}
for which
we refer to the $n$-qubit state coupled to the ancillary state
$| 1 \rangle \otimes | 0 \rangle$
as the success state.
By setting the real-time step to $\Delta t_1$ defined in
Eq.~(\ref{eigenstate_canceller:optimal_dt}),
the unnormalized success state 
from $\mathcal{C}_{\mathrm{filt2}}^{(\widetilde{E}_0)}$
is written as
(see Appendix \ref{sec:derivation_for_filtration})
\begin{gather}
    \sum_k
        c_k
        \sin^2
        \frac{\pi (E_k - \widetilde{E}_0) }
        {2 (\widetilde{E}_1 - \widetilde{E}_0 )}
    | \phi_k \rangle
    ,
    \label{eigenstate_canceller:success_state_for_filt2}
\end{gather}
where the amplitude of ground state is $\mathcal{O} (\delta E_0^2)$
in contrast to the first-order filtration circuit.
The approximate upper bound of the allowed total imaginary-time lapse is given by
\begin{align}
    \tau_{\mathrm{ub 2}}
    \approx
        \frac{1}{E_1 - E_0}
        \left(
            \ln \frac{16 \varepsilon}{\pi^4}
            +
            2
            \ln
            \left| \frac{c_1}{c_0} \right|
            +
            4
            \ln
            \frac{1}{| \overline{\delta} E_0 |}
        \right)
    ,
    \label{filtration_ub2_approx}
\end{align}
larger than $\tau_{\mathrm{ub 1}}$
in Eq.~(\ref{filtration_ub1_approx}).

The first- and second-order filtration techniques 
for the ground state and the first excited state
described above are also applicable to other combinations of the energy eigenstates.

\section{Results}
\label{sec:results}

\subsection{Ground state of an electron in a harmonic potential}
\label{sec:results:harmonic}

Here we perform simulations of PITE for the Fock--Darwin system \cite{bib:5625, bib:5626, bib:5627},
which models the orbital motion of a single electron confined to a plane feeling a harmonic potential and a uniform magnetic field perpendicular to the plane.

\subsubsection{Setup}

We assume that the electron is on the $xy$ plane
and, by defining $X \equiv x - L/2$ and $Y \equiv y - L/2,$ 
the potential is given by $V (x, y) = m \omega_0^2 (X^2 + Y^2)/2.$
Recalling that an electron under a uniform magnetic field in a GaAs quantum dot is often studied (see, e.g., Refs.~\cite{bib:5633, bib:5629, bib:5635, bib:5637}),
we adopt the effective electron mass $m = 0.067 m_e$,
where $m_e$ is the bare electron mass,
and the confining strength $\omega_0 = 4$ meV.
We set $B = 5$ T
for the shifted vector potential
$\boldsymbol{A} (\boldsymbol{r}) = (x - L/2) B \boldsymbol{e}_y$
and ignore the spin degree of freedom for simplicity.

We encoded the real-space wave function of the electron
by using six qubits for each of the $x$ and $y$ directions. 
The numerically exact energy eigenstates were obtained by diagonalizing the Hamiltonian matrix in the position representation (see Appendix \ref{sec:Hamiltonian_mat_of_1p}).
The size $L = 120$ nm of the simulation cell was confirmed to be large enough to get the converged low-energy spectra
(see Appendix \ref{sec:spectra_for_harmonic}).
The CNOT gate count for the PITE circuit is provided in
Appendix~\ref{sec:gate_counts}.

\subsubsection{Ground state from Gaussian initial state}

By using the normalized Gaussian wave function
\begin{gather}
    g (x, y; x_{\mathrm{c}}, w)
    \propto
    \exp
    \left(
        -\frac{(X - x_{\mathrm{c}})^2 + Y^2}{w^2}
    \right)
    ,
\end{gather}
where $x_{\mathrm{c}}$ specifies the center and $w$ is the width,
we defined the initial state
$
\psi_{\mathrm{init}} (x, y)
\equiv
g (x, y; 0, 20 \ \mathrm{nm})
$
centered at the potential bottom.
We performed the PITE circuit simulations starting from this initial state
for some combinations of the splitting pattern and the amount $\Delta \tau$ of each step.
We used $m_0 = 0.9$ in the ITE operator $\mathcal{M}$ for all the simulations.
The weight of ground state in the input state at each PITE step
for constant-$\Delta \tau$ cases
is plotted in the left panel of Fig.~\ref{fig:single_well_weights}(a).
The total success probability at each step is also plotted.
In the cases for $\Delta \tau = 0.02$ meV$^{-1},$
we can see that
the weights for the two splitting patterns do not differ much from each other due to the small $\Delta \tau.$
Those for $\Delta \tau = 0.05$ meV$^{-1},$
on the other hand, the PITE steps exhibited unstable behavior
due to the large step.
The oscillatory behavior that may occur when approaching the ground state is a consequence of the approximation using the RTE operators instead of the exact ITE operator (see also Nishi et al.~\cite{Nishi_PITE_cost}).
In addition, the Suzuki-Trotter splitting within each RTE operator can excite the trial state.
These two facts make it difficult to find the optimal choice of $\Delta \tau.$

Given the observations above,
we performed simulations
for a variable imaginary-time step defined in
Eq.~(\ref{def_variable_dtau})
with
$\Delta \tau_{\mathrm{min}} = 0.02$ meV$^{-1}$,
$\Delta \tau_{\mathrm{max}} = 0.05$ meV$^{-1}$, and
$\kappa = 5.$
The results are plotted in the right panel of
Fig.~\ref{fig:single_well_weights}(a).
The $TVT$ splitting gave the better convergence than the $TV$ splitting and the total success probability for the former is higher than that of the latter.
In addition, the convergence for the $TVT$ splitting is better than in the constant-$\Delta \tau$ cases shown in the left panel.
The reason why the variable-$\Delta \tau$ cases can reach the ground state in contrast to the cases of constant large $\Delta \tau$
is that 
the high-energy states having significant weights at the early PITE steps are
rotated rapidly by a large $\Delta \tau$ in the RTE operator,
rendering the approximate ITE inaccurate.

\subsubsection{Ground state from exponential initial state}

\begin{figure*}
\begin{center}
\includegraphics[width=16cm]{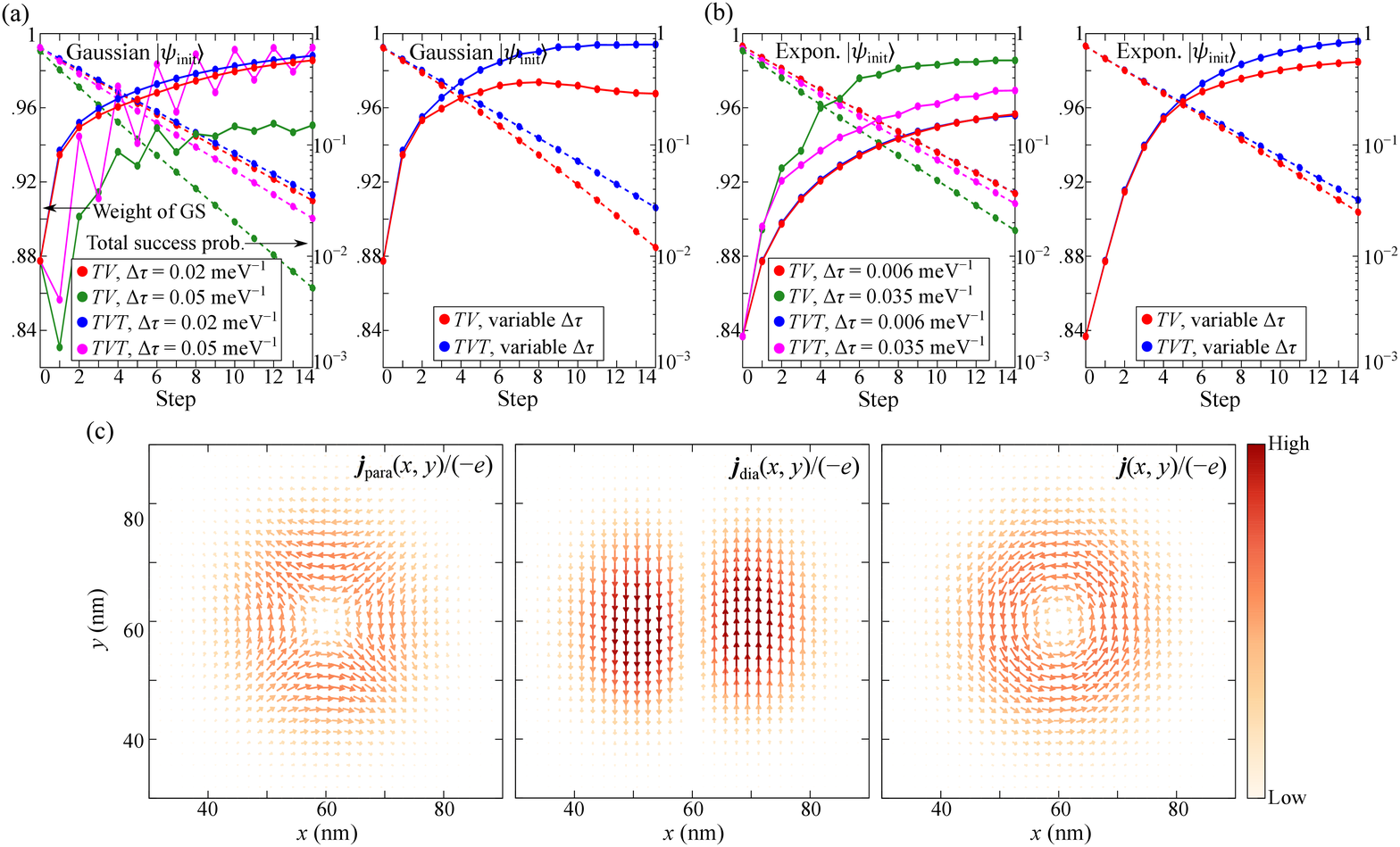}
\end{center}
\caption{
(a)
Circles connected by solid lines in the left panel show
the weight of ground state (GS) in the input state at each PITE step
with a constant $\Delta \tau$ for a Gaussian initial state.
Circles connected by dashed lines in the left panel show
the total success probability at each step.
The right panel shows similar plots for 
$\Delta \tau_{\mathrm{min}} = 0.02$ meV$^{-1}$,
$\Delta \tau_{\mathrm{max}} = 0.05$ meV$^{-1}$, and
$\kappa = 5.$
(b)
Results for an initial state having an exponential shape.
The left panel is for constant-$\Delta \tau$ cases,
while the right panel is for
$\Delta \tau_{\mathrm{min}} = 0.006$ meV$^{-1}$,
$\Delta \tau_{\mathrm{max}} = 0.035$ meV$^{-1}$, and
$\kappa = 5.$
(c) 
The paramagnetic, diamagnetic,
and total probability current density
calculated from the measurement-based formulae for the final state for
the $TVT$ splitting with the variable $\Delta \tau.$
}
\label{fig:single_well_weights}
\end{figure*}

Since an efficient way for preparing an exponentially decaying state is known \cite{bib:5645},
we also performed simulations by using an exponential initial state
$\psi_{\mathrm{init}} (x, y) \propto \exp (-(|X| + |Y|)/d),$
where $d = 15$ nm is the decay length.
The weight of ground state in the input state at each PITE step
for constant-$\Delta \tau$ cases
is plotted in the left panel of Fig.~\ref{fig:single_well_weights}(b).
In the cases for $\Delta \tau = 0.006$ meV$^{-1}$,
the convergence is slow for both splitting patterns.
For $\Delta \tau = 0.035$ meV$^{-1}$,
the convergence is faster.
The $TVT$ splitting, however, exhibits the slower convergence than the $TV$ splitting.
The right panel of the figure shows the results
for a variable step with
$\Delta \tau_{\mathrm{min}} = 0.006$ meV$^{-1}$,
$\Delta \tau_{\mathrm{max}} = 0.035$ meV$^{-1}$, and
$\kappa = 5,$
where the $TVT$ splitting gives rise to more rapid convergence than the $TV$ splitting.
The convergence for the $TVT$ splitting is better than in the constant-$\Delta \tau$ cases shown in the left panel,
similarly to the cases with the Gaussian initial state.

The results obtained in the simulations above suggest that
it is a good choice for adopting the combination of $TVT$ splitting and a variable $\Delta \tau$ for achieving rapid convergence for a generic system.

We also simulated the construction of the probability current density,
defined as the electric current density divided by $-e,$
by using the measurement-based formulae in
Eqs.~(\ref{electron_dens_from_meas}) and
(\ref{current_dens_paramag_x_from_meas}).
Figure \ref{fig:single_well_weights}(c)
shows the results for the final state obtained for
the $TVT$ splitting with the variable $\Delta \tau.$
The calculated total current density correctly exhibits the counterclockwise motion of the electron around the potential bottom,
as understood from the classical picture via the Lorentz force \cite{bib:4010, bib:5627}.

\subsection{Excited states of an electron in a double-well potential}
\label{sec:results:double_well}

Here we perform simulations of PITE for an electron confined to a double-well potential that mimics a GaAs double quantum dot
adopted by Abolfath et al.~\cite{bib:5635, bib:5637}

\subsubsection{Setup}

The potential is given by
\begin{gather}
    V (x, y)
    =
        V_0
        \exp \left( - \frac{(X + a)^2 + Y^2}{\Delta^2} \right)
    \nonumber \\
        +
        V_0
        \exp \left( - \frac{(X - a)^2 + Y^2}{\Delta^2} \right)
        +
        V_p
        \exp \left( - \frac{X^2}{\Delta_x^2} - \frac{Y^2}{\Delta_y^2} \right)
        ,
    \label{def_double_well_pot}
\end{gather}
where $X \equiv x - L/2$ and $Y \equiv y - L/2.$  
$V_0 = -59.3$ meV represents the left and right dot potentials.
$V_p = 41.51$ meV represents the barrier generated by the plunger gate.
$a = 2$ nm specifies the location of left and right dots
having widths $\Delta = 24.48$ nm for the symmetric shapes,
while the barrier has $\Delta_x = 2.94$ nm and $\Delta_y = 24.48$ nm
for the asymmetric shape.
We introduce an external magnetic field $B = 3$ T along the $z$ direction for the shifted vector potential
and adopt the effective electron mass
similarly to the case of a harmonic potential described above.
We ignore the spin degree of freedom for simplicity.

We encoded the real-space wave function of the electron
by using six qubits for each of the $x$ and $y$ directions. 
The size $L = 120$ nm of the simulation cell was confirmed to be large enough to get the converged low-energy spectra
(see Appendix \ref{sec:spectra_for_double_well}).
We adopted the $TVT$ splitting and $m_0 = 0.9$ for all the PITE simulations described below.
The CNOT gate count for the PITE circuit is provided in
Appendix~\ref{sec:gate_counts}.

\subsubsection{Lowest-energy states within subspaces}

Since the target system subject to the magnetic field is invariant under inversion
$(X \rightarrow -X, \ Y \rightarrow -Y),$
each energy eigenstate has its own positive or negative parity.
We found,
by looking into the numerically obtained energy eigenstates
$| \phi_\nu \rangle \ (\nu = 0, 1, \dots)$,
that
$
| \phi_0 \rangle,
| \phi_2 \rangle,
| \phi_5 \rangle,
| \phi_6 \rangle$, and
$| \phi_8 \rangle$
have positive parity among the lowest ten states,
while the other five have negative parity.

\begin{figure*}
\begin{center}
\includegraphics[width=18cm]{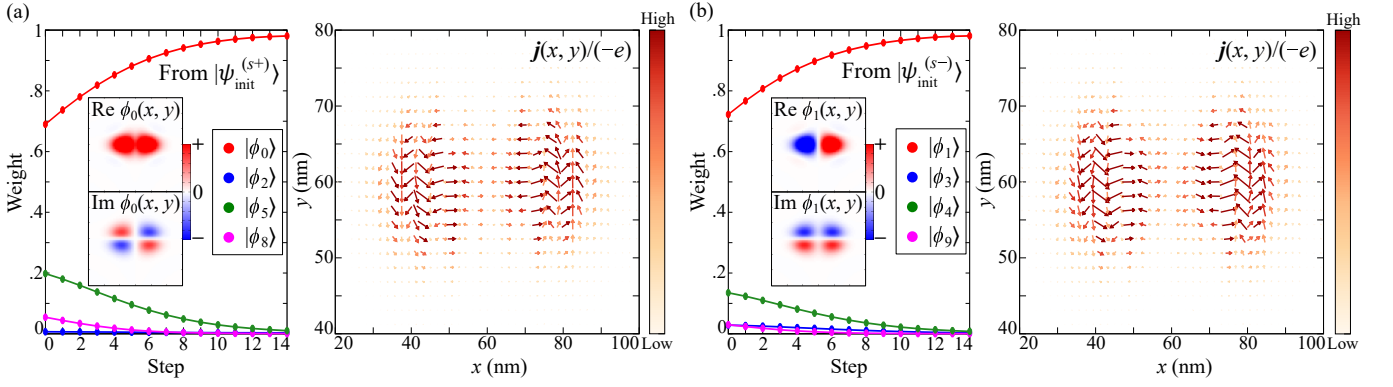}
\end{center}
\caption{
(a)
Left panel shows the weights of the energy eigenstates contained in the input state at each PITE step
starting from $| \psi_{\mathrm{init}}^{(s+)} \rangle.$
The inset draws the ground state in real space.
The right panel shows the total probability current density of the final state.
(b)
Similar plot for $| \psi_{\mathrm{init}}^{(s-)} \rangle.$
The inset draws the first excited state in real space.
}
\label{fig:double_well_weights}
\end{figure*}

To obtain the ground state $| \phi_0 \rangle$,
we defined the initial state as a bonding $s$-type orbital
\begin{gather}
    \psi_{\mathrm{init}}^{(s+)} (x, y)
    \propto
        g (x, y; a, w)
        +
        g (x, y; -a, w)
    \label{double_well_def_init_bonding_s}
\end{gather}
with the widths $w = 11$ nm of Gaussian functions,
which has positive parity.
We performed a PITE circuit simulation starting from this initial guess
by using variable imaginary-time step with
$\Delta \tau_{\mathrm{min}} = 0.004$ meV$^{-1}$,
$\Delta \tau_{\mathrm{max}} = 0.008$ meV$^{-1}$, and
$\kappa = 10.$
The iterations reached the convergence,
as shown in the left panel of Fig.~\ref{fig:double_well_weights}(a),
where the parity has been correctly preserved throughout the iterations to the final state
$| \psi_{\mathrm{fin}}^{(s+)} \rangle.$
We also performed a simulation for obtaining the first excited state
$| \phi_1 \rangle$,
which is the lowest-energy state within the negative-parity subspace.
We defined the initial state as an anti-bonding $s$-type orbital
\begin{gather}
    \psi_{\mathrm{init}}^{(s-)} (x, y)
    \propto
        g (x, y; a, w)
        -
        g (x, y; -a, w)
        ,
    \label{double_well_def_init_anti_bonding_s}
\end{gather}
which has negative parity.
As seen in the results shown in the left panel of Fig.~\ref{fig:double_well_weights}(b),
the iterations continued by preserving the parity to reach the final state
$| \psi_{\mathrm{fin}}^{(s-)} \rangle.$

We calculated the probability current densities of
$| \psi_{\mathrm{fin}}^{(s+)} \rangle$
and
$| \psi_{\mathrm{fin}}^{(s-)} \rangle$
from the measurement-based formulae,
as shown in the right panels of
Figs.~\ref{fig:double_well_weights}(a) and (b),
respectively.
In contrast to the opposite parity of the two states,
the motions of the electron in them
basically coincide with each other.
That is, the electron exhibits counterclockwise motion around each potential minimum.

\subsubsection{Filtration for obtaining excited states}

\begin{figure}
\begin{center}
\includegraphics[width=8.5cm]{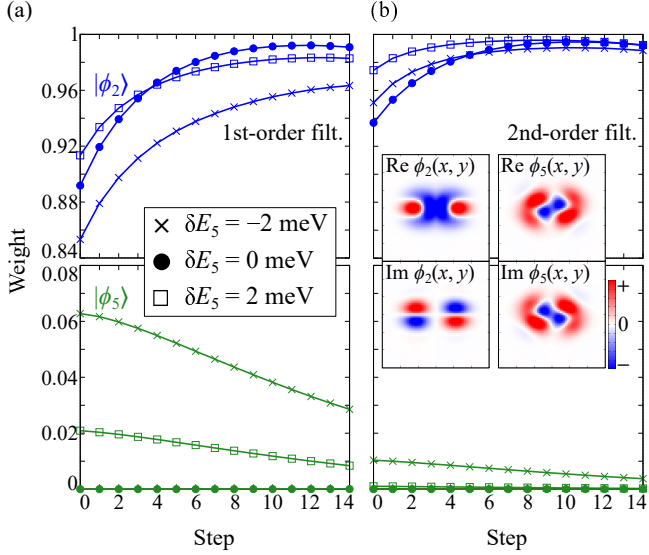}
\end{center}
\caption{
(a)
Weights of $| \phi_2 \rangle$ and $| \phi_5 \rangle$ in the input state
at each PITE step for obtaining $| \phi_2 \rangle.$
The PITE steps started after applying
the first-order filtration circuits
with an error $\delta E_5$
to $| \psi_{\mathrm{init}}^{(p_x +)} \rangle$
for removing $| \phi_0 \rangle$ and $| \phi_5 \rangle.$
(b)
Similar plots for the second-order filtration circuits.
The inset draws $| \phi_2 \rangle$ and $| \phi_5 \rangle$ in real space.
}
\label{fig:weights_for_phi2}
\end{figure}

To obtain the second excited state $| \phi_2 \rangle,$
we defined the initial state as a $\sigma$ bonding $p_x$-type orbital
\begin{gather}
    \psi_{\mathrm{init}}^{(p_x +)} (x, y)
    \propto
        g \left( x, y; \frac{3 a}{2}, w \right)
        +
        g \left( x, y; -\frac{3 a}{2}, w \right)
    \nonumber \\
        -
        \frac{5}{2}
        g ( x, y; 0, w )
        ,
    \label{double_well_def_bonding_px}
\end{gather}
which has positive parity.
We found that this initial guess contains
a significant weight $0.45$ of $| \phi_2 \rangle$
and a non-negligible weight $0.22$ of $| \phi_5 \rangle,$
as inferred from their shapes depicted in the insets of
Fig.~\ref{fig:weights_for_phi2}.
Therefore,
in order for the PITE steps to converge to $| \phi_2 \rangle$ rapidly,
not only $| \phi_0 \rangle$ but also $| \phi_5 \rangle$
have to be removed from the initial state.
To this end,
we applied the first-order filtration circuits
$\mathcal{C}_{\mathrm{filt1}}^{(\widetilde{E}_0)}$ and 
$\mathcal{C}_{\mathrm{filt1}}^{(\widetilde{E}_5)},$
described in Sect.~\ref{sec:filtration},
for which we assumed the estimated energy eigenvalues of 
$| \phi_0 \rangle$ and $| \phi_2 \rangle$ have no error
$(\delta E_0 = \delta E_2 = 0)$ and that of $| \phi_5 \rangle$ can have an error.
The results of PITE steps starting from the filtered input state are shown in Fig.~\ref{fig:weights_for_phi2}(a)
for three values of the error $\delta E_5.$
We used the variable imaginary-time step with
$\Delta \tau_{\mathrm{min}} = 0.003$ meV$^{-1}$,
$\Delta \tau_{\mathrm{max}} = 0.005$ meV$^{-1}$, and
$\kappa = 10.$
The results for the input state with the second-order filtration circuits
$\mathcal{C}_{\mathrm{filt2}}^{(\widetilde{E}_0)}$ and 
$\mathcal{C}_{\mathrm{filt2}}^{(\widetilde{E}_5)}$
are also shown in Fig.~\ref{fig:weights_for_phi2}(b).
It is seen that the weight of $| \phi_2 \rangle$ can decrease slightly after the tenth step despite the monotonic decrease in the weight of $| \phi_5 \rangle.$
This behavior implies that further application of filtration for removing energy eigenstates other than $| \phi_5 \rangle$ on the way to $| \phi_2 \rangle$ is needed.

\begin{figure}
\begin{center}
\includegraphics[width=6.5cm]{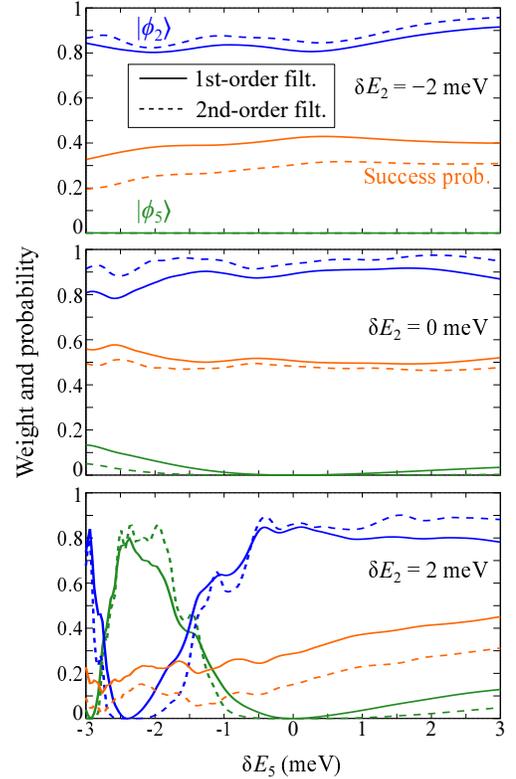}
\end{center}
\caption{
Weights of $| \phi_2 \rangle$ and $| \phi_5 \rangle$
in the initial state after applying
the filtration circuits
to $| \psi_{\mathrm{init}}^{(p_x +)} \rangle$
for removing $| \phi_0 \rangle$ and $| \phi_5 \rangle$
are plotted
as functions of the errors $\delta E_2$ and $\delta E_5.$
The success probabilities of filtration are also plotted.
}
\label{fig:double_well_filt}
\end{figure}

The weights of $| \phi_2 \rangle$ and $| \phi_5 \rangle$
in the initial state after applying
the filtration circuits
to $| \psi_{\mathrm{init}}^{(p_x +)} \rangle$
for removing $| \phi_0 \rangle$ and $| \phi_5 \rangle$
are plotted
in Fig.~\ref{fig:double_well_filt}
as functions of the errors $\delta E_2$ and $\delta E_5$ with $\delta E_0 = 0.$
The success probabilities of filtration are also plotted.
It is observed that these weights are asymmetric with respect to the signs of $\delta E_2$ and $\delta E_5.$
The weights of $| \phi_5 \rangle$ around $\delta E_5 = 0$
is more suppressed by the second-order filtration
than by the first-order one, as expected.
The reason for the success probability for the second-order filtration
being lower than that for the first-order may be simply that
the former distributes the weights in the input to the four possible ancillary states,
while the latter does that to the two possible ancillary states.

\section{Conclusions}
\label{sec:conclusions}

We proposed a method to perform FQE calculation for an electronic system under a uniform magnetic field.
The magnetic-phase gate originating from the decoupled evolution operators was found to lead to efficient implementation of the PITE circuit.
Resource estimation revealed that
introducing a magnetic field into an interacting electronic system does not significantly raise the entire computational cost.

We performed FQE simulations for a confined electron under a uniform magnetic field,
and confirmed that 
our method correctly provides the ground state,
for which the measurement-based formulae for the electric current density was found to work.
In addition, the new method in combination with the first- and second-order filtration circuits produced the desired excited states.

The methods proposed in this study allows one to use the FQE framework to obtain the lowest-energy states of a generic molecular or model system subjected to a uniform magnetic field.
The techniques for incorporating the magnetic field are also applicable to quantum simulations of real-time dynamics in the first-quantized form. 
These facts will extend  the practical application of quantum computation to analyze static and dynamical properties of interacting electronic systems.

\begin{acknowledgments}
This work was supported by MEXT as ``Program for Promoting Researches on the Supercomputer Fugaku'' (JPMXP1020200205) and JSPS KAKENHI as ``Grant-in-Aid for Scientific Research(A)'' Grant Number 21H04553. The computation in this work has been done using (supercomputer Fugaku provided by the RIKEN Center for Computational Science/Supercomputer Center at the Institute for Solid State Physics in the University of Tokyo).
\end{acknowledgments}

\begin{widetext}

\appendix

\section{Derivation of Eq.~(\ref{PITE_with_mag_fields:impl_of_U_kin})}
\label{sec:derivation_of_kin_phase}

For each direction, 
the action of kinetic-phase gate
on the computational basis $| j \rangle_n \ (j = 0, 1, \dots, N - 1)$ is
\begin{align}
    U_{\mathrm{kin}} (\Delta t)
    | j \rangle_n
    &=
        \exp
        \left(
            -i \frac{(\widetilde{j} \Delta p)^2}{2 m}
            \Delta t
        \right)
        | j \rangle_n
    \nonumber \\
    &=
        e^{-i e_{\mathrm{kin}} j^2 \Delta t}
        e^{i N e_{\mathrm{kin}} j \Delta t}
        e^{-i N^2 e_{\mathrm{kin}} \Delta t/4}
        | j \rangle_n
        .
    \label{PITE_with_mag_fields:action_of_U_kin}
\end{align}
By using the binary digits for
$j =  \sum_{\ell = 0}^{n - 1} 2^\ell j_{\ell}$,
the contribution from the first-order exponent
on the right-hand side of in Eq.~(\ref{PITE_with_mag_fields:action_of_U_kin})
is written as
\begin{align}
    e^{i N e_{\mathrm{kin}} j \Delta t}
    | j \rangle_n
    &=
        \prod_{\ell = 0}^{n - 1}
            \exp (i 2^\ell j_\ell N e_{\mathrm{kin}} \Delta t)
        | j \rangle_n
    \nonumber \\
    &=
        \left(
            Z_{\mathrm{kin}} (2^{n - 1} N)
            | j_{n - 1} \rangle
        \right)
        \otimes
        \cdots
        \otimes
        \left(
            Z_{\mathrm{kin}} (2^1 N)
            | j_1 \rangle
        \right)
        \otimes
        \left(
            Z_{\mathrm{kin}} (2^0 N)
            | j_0 \rangle
        \right)
    \nonumber \\
    &=
            \prod_{\ell = 0}^{n - 1}
            \overbrace{
                Z_{\mathrm{kin}}
                (2^\ell N)
            }^{\mathrm{On} \ \ell \mathrm{th \ qubit} }
        | j \rangle_n
    ,
    \label{PITE_with_mag_fields:action_of_U_kin_1st_order}
\end{align}
where we introduced the single-qubit phase gate $Z_{\mathrm{kin}}$
defined in Eq.~(\ref{def_phase_kin_1q}).
The contribution from the second-order exponent is written as
\begin{align}
    e^{-i e_{\mathrm{kin}} j^2 \Delta t}
    | j \rangle_n
    &=
        \prod_{\ell = 0}^{n - 1}
        \prod_{\ell' = 0}^{n - 1}
            \exp
            (
                -i
                2^{\ell + \ell'}
                j_\ell j_{\ell'}
                e_{\mathrm{kin}} \Delta t
            )
        | j \rangle_n
    \nonumber \\
    &=
        \left(
            \prod_{\ell = 0}^{n - 1}
            \exp
            (
                -i
                2^{2 \ell}
                j_\ell
                e_{\mathrm{kin}} \Delta t
            )
        \right)
            \prod_{\ell = 0}^{n - 1}
            \prod_{\ell' \ne \ell}^{n - 1}
                \exp
                (
                    -i
                    2^{\ell + \ell'}
                    j_\ell j_{\ell'}
                    e_{\mathrm{kin}} \Delta t
                )
        | j \rangle_n
    \nonumber \\
    &=
        \left(
            Z_{\mathrm{kin}} (-2^{2 n - 2})
            \otimes
            \cdots
            \otimes
            Z_{\mathrm{kin}} (-2^2)
            \otimes
            Z_{\mathrm{kin}} (-2^0) 
        \right)
            \prod_{\ell = 0}^{n - 1}
            \prod_{\ell' \ne \ell}^{n - 1}
                \mathrm{C}_{\ell, \ell'}
                Z_{\mathrm{kin}} (-2^{\ell + \ell'})
        | j \rangle_n
    \nonumber \\
    &=
        \left(
            Z_{\mathrm{kin}} (-2^{2 n - 2})
            \otimes
            \cdots
            \otimes
            Z_{\mathrm{kin}} (-2^2)
            \otimes
            Z_{\mathrm{kin}} (-2^0)
        \right)
            \prod_{\ell = 0}^{n - 1}
            \prod_{\ell' < \ell}
                \mathrm{C}_{\ell, \ell'}
                Z_{\mathrm{kin}} (-2^{\ell + \ell' + 1})
        | j \rangle_n
        ,
    \label{PITE_with_mag_fields:action_of_U_kin_2nd_order}
\end{align}
where the two-qubit gate
$\mathrm{C}_{\ell, \ell'} Z_{\mathrm{kin}}$
acts on the $\ell'$th qubit as $Z_{\mathrm{kin}}$ 
controlled by the $\ell$th qubit.
These gates clearly commutes with each other. 
We used the relation
$\mathrm{C}_{\ell, \ell'} Z_{\mathrm{kin}}
=
\mathrm{C}_{\ell', \ell} Z_{\mathrm{kin}}
$
to get the last equality in 
Eq.~(\ref{PITE_with_mag_fields:action_of_U_kin_2nd_order}).

Substitution of Eqs.~(\ref{PITE_with_mag_fields:action_of_U_kin_1st_order})
and (\ref{PITE_with_mag_fields:action_of_U_kin_2nd_order})
into Eq.~(\ref{PITE_with_mag_fields:action_of_U_kin})
leads to Eq.~(\ref{PITE_with_mag_fields:impl_of_U_kin}).

\section{Derivation of Eq.~(\ref{PITE_with_mag_fields:impl_of_U_mag})}
\label{sec:derivation_of_mag_phase}

Let us consider the position eigenstate
$| x^{(k)} \rangle \otimes | y^{(k')} \rangle$
specified by the discretized coordinates
$x^{(k)} = k \Delta x$
and
$y^{(k')} = k' \Delta x.$
By using the binary digits for
$k = \sum_{\ell = 0}^{n - 1} 2^\ell k_{\ell}$
and those for $k'$ similarly,
the action of magnetic-phase gate on the position eigenstate is written as
\begin{align}
    U_{\mathrm{mag}}
    | x^{(k)} \rangle \otimes | y^{(k')} \rangle
    &=
        \exp
        \left( i \mu x^{(k)} y^{(k')} \right)
        | x^{(k)} \rangle \otimes | y^{(k')} \rangle
    \nonumber \\
    &=
        \exp
        \left(
            i \mu
            \Delta x^2
            \sum_{\ell = 0}^{n - 1}
            \sum_{\ell' = 0}^{n - 1}
            2^{\ell + \ell'}
            k_{\ell}
            k'_{\ell'}
        \right)
        | x^{(k)} \rangle \otimes | y^{(k')} \rangle
    \nonumber \\
    &=
        \prod_{\ell = 0}^{n - 1}
        \prod_{\ell' = 0}^{n - 1}
        \exp
        \left(
            i
            2^{\ell + \ell'}
            k_{\ell}
            k'_{\ell'}
            \mu
            \Delta x^2
        \right)
        | x^{(k)} \rangle \otimes | y^{(k')} \rangle
    \nonumber \\
    &=
        \prod_{\ell = 0}^{n - 1}
        \prod_{\ell' = 0}^{n - 1}
        \mathrm{C}_{\ell, \ell'}
        Z_{\mathrm{mag}}^{(\ell + \ell')}
        | x^{(k)} \rangle \otimes | y^{(k')} \rangle
        ,
    \label{PITE_with_mag_fields:action_of_U_mag}
\end{align}
where we introduced the single-qubit phase gate $Z_{\mathrm{mag}}$
defined in Eq.~(\ref{def_phase_mag_1q}).
The two-qubit gate
$\mathrm{C}_{\ell, \ell'} Z_{\mathrm{mag}}^{(\ell + \ell')}$
represents the $Z_{\mathrm{mag}}^{(\ell + \ell')}$ gate
on the $\ell'$th qubit for the $y$ position coordinate
controlled by the $\ell$th qubit for the $x$ position coordinate.
From the fact that these gates commute with each other
and Eq.~(\ref{PITE_with_mag_fields:action_of_U_mag}),
it is now clear that $U_{\mathrm{mag}}$ can be written as 
Eq.~(\ref{PITE_with_mag_fields:impl_of_U_mag}).

\section{Calculation of derivative distributions via repeated measurements}
\label{sec:fdfc_distr_from_measurements}

Here we consider a differentiable complex function $f$ of $m$ continuous real variables $x_0, \dots, x_{m - 1}.$
We denote them by $\boldsymbol{x}$ collectively.
We assume that we do not know the functional form of $f.$
For a given $r,$
we want to calculate the distributions
\begin{align}
    g_{r_0, r_1, \dots, r_{m - 1}}^{(r_0 + r_1 + \dots + r_{m - 1})}
    (\boldsymbol{x})
    \equiv
        f (\boldsymbol{x})
        \frac{1}{r_0 ! r_1 ! \cdots r_{m - 1} !}
        \frac{\partial^{r_0 + r_1 + \dots + r_{m - 1}} f (\boldsymbol{x})^*}{\partial x_0^{r_0} \partial x_1^{r_1} \cdots \partial x_{m - 1}^{r_{m - 1}}}
        ,
    \label{deriv_of_func_as_many_qubits:def_g}
\end{align}
which we call the derivative distributions,
as functions of $\boldsymbol{x}$
for all the possible combinations
$(r_0, \dots, r_{m - 1})$
satisfying $r_0 + \cdots + r_{m - 1} \leq r.$
We describe below a generic procedure for calculating them via repeated measurements for qubits which encode $f.$

\subsection{Linear equations based on finite differences}

By performing the multivariate Taylor expansion of $f$ around a point $\boldsymbol{x},$
we can write
\begin{align}
    f (\boldsymbol{x} + \boldsymbol{h})
    =
        \sum_{\ell = 0}^\infty
        \sum_{r_0 + \cdots + r_{m - 1} = \ell}
        \frac{1}{r_0 ! \cdots r_{m - 1} !}
        \frac{\partial^\ell f (\boldsymbol{x})}
            {\partial x_0^{r_0} \dots \partial x_{m - 1}^{r_{m - 1}}}
        h_0^{r_0}
        \cdots
        h_{m - 1}^{r_{m - 1}}
    \label{deriv_of_func_as_many_qubits:Taylor_expansion_of_f}
\end{align}
for an arbitrary $m$-dimensional vector $\boldsymbol{h}.$
This expression and 
Eq.~(\ref{deriv_of_func_as_many_qubits:def_g})
lead to
\begin{align}
    f (\boldsymbol{x})
    f (\boldsymbol{x} + \boldsymbol{h})^*
    &=
        \sum_{\ell = 0}^\infty
        \sum_{(r_0, \dots, r_{m - 1}) \in \mathcal{D}(\ell)}
        g_{r_0, \dots, r_{m - 1}}^{(\ell)}
        (\boldsymbol{x})
        h_0^{r_0}
        \cdots
        h_{m - 1}^{r_{m - 1}}
        ,
    \label{deriv_of_func_as_many_qubits:fx_fxhc_exact}
\end{align}
where
$\mathcal{D}(\ell)$ is the set of
the combinations $(r_0, \dots, r_{m - 1})$ of nonnegative integers
that satisfy $r_0 + \cdots + r_{m - 1} = \ell$
for the $\ell$th-order derivatives.
We can easily understand that the number of elements in
$\mathcal{D}(\ell)$ is given by
\begin{align}
    |\mathcal{D}(\ell)|
    =
        \frac{(\ell + m - 1)!}{\ell ! (m - 1)!}
        .
    \label{deriv_of_func_as_many_qubits:num_of_integers_in_D}
\end{align}
We can see Eq.~(\ref{deriv_of_func_as_many_qubits:fx_fxhc_exact}) as a linear equation for an infinite number of  unknown distributions
$g_{r_0, \dots, r_{m - 1}}^{(\ell)}.$

To eliminate the contributions on the order of $h^{r + 1},$
we define
\begin{align}
    G^{(r)} (\boldsymbol{x}, \boldsymbol{h})
    &\equiv
        \frac{
            f (\boldsymbol{x})
            f (\boldsymbol{x} + \boldsymbol{h})^*
            +
            (-1)^r
            f (\boldsymbol{x})
            f (\boldsymbol{x} - \boldsymbol{h})^*
        }{2}
    \label{deriv_of_func_as_many_qubits:def_G}
\end{align}
for an integer $r.$
Substituting 
Eq.~(\ref{deriv_of_func_as_many_qubits:Taylor_expansion_of_f})
into this definition,
we obtain
\begin{align}
    G^{(r)} (\boldsymbol{x}, \boldsymbol{h})
    &=
        \begin{cases}
            \sum_{\mathrm{even} \ \ell}^r
            \sum_{(r_0, \dots, r_{m - 1}) \in \mathcal{D}(\ell)}
            g_{r_0, \dots, r_{m - 1}}^{(\ell)}
            (\boldsymbol{x})
            h_0^{r_0}
            \cdots
            h_{m - 1}^{r_{m - 1}}
            +
            \mathcal{O} (h^{r + 2})
            &
            \mathrm{even} \ r
            \\
            \sum_{\mathrm{odd} \ \ell}^r
            \sum_{(r_0, \dots, r_{m - 1}) \in \mathcal{D}(\ell)}
            g_{r_0, \dots, r_{m - 1}}^{(\ell)}
            (\boldsymbol{x})
            h_0^{r_0}
            \cdots
            h_{m - 1}^{r_{m - 1}}
            +
            \mathcal{O} (h^{r + 2})
            &
            \mathrm{odd} \ r
        \end{cases}
    .
    \label{deriv_of_func_as_many_qubits:lin_eq_for_even_and_odd}
\end{align}
We omit the terms whose orders are higher than $h^r$
in this expression 
to construct a linear equation for the finite number of unknowns.
The number of unknowns is given by
\begin{align}
    N^{(r)}
    =
    \begin{cases}
        \sum_{\mathrm{even} \ \ell}^r
        | \mathcal{D}(\ell) |
        &
        \mathrm{even} \ r
        \\                
        \sum_{\mathrm{odd} \ \ell}^r
        | \mathcal{D}(\ell) |
        &
        \mathrm{odd} \ r
    \end{cases}
    .
    \label{deriv_of_func_as_many_qubits:num_of_unknowns_in_lin_eq}
\end{align}
The linear equation for an even (odd) $r$ provides an approximate solution with an error on the order of $h^{r + 2}$ ($h^{r + 1}$). 
We need $N^{(r)}$ independent conditions for solving the linear equation.
In practice, we adopt some small vectors
$\boldsymbol{h}_d \ (d = 0, \dots, N^{(r)} - 1)$
and calculate 
$G^{(r)} (\boldsymbol{x}, \boldsymbol{h}_d)$
via the definition in 
Eq.~(\ref{deriv_of_func_as_many_qubits:def_G})
to construct the linear equation in
Eq.~(\ref{deriv_of_func_as_many_qubits:lin_eq_for_even_and_odd})
to be solved on a classical computer.

\subsection{Circuits for calculation of $G^{(r)}$}

If a many-qubit system which encodes $f$ is available,
it is possible to obtain the derivative distributions by using the circuits explained below.

We assume that we have $n$ qubits for each of the $m$ variables and assign each computational basis
$| j \rangle_n \ (j = 0, \dots, 2^n - 1)$
to equidistant grid points on a range $[0, L]$:
$x^{(j)} = j \Delta x,$
where $\Delta x \equiv L/N.$
We assume further that
the unknown function $f$ is available
as a normalized $m n$-qubit state
\begin{align}
    | f \rangle
    =
        \sum_{j_0, \dots, j_{m - 1}}^{N - 1}
        f (x^{(j_0)}, \dots, x^{(j_{m - 1})})
        | j_0 \rangle_n
        \otimes
        \cdots
        \otimes
        | j_{m - 1} \rangle_n
        ,
\end{align}
where $N \equiv 2^n.$
The probability for finding $| \boldsymbol{x} \rangle$ 
in an $m n$-qubit measurement on $| f \rangle$ is
$\mathbb{P}_{\boldsymbol{x}} = |f (\boldsymbol{x})|^2.$
We want to obtain the derivative distributions,
defined in
Eq.~(\ref{deriv_of_func_as_many_qubits:def_g}),
at the grid points.
To this end, we have to calculate
$G^{(r)} (\boldsymbol{x}, \boldsymbol{h})$
for an arbitrary displacement vector $\boldsymbol{h}$
for constructing the linear equation.

From an $n$-qubit phase gate
$U_{\mathrm{shift}} (d)$ for an integer $d$ defined via
$
| j \rangle_n
\xmapsto{}
\exp(-i 2 \pi d j/N )
| j \rangle_n
\
(j = 0, \dots, N - 1)
,
$
we define
\begin{align}
    \mathcal{U} (d)
    \equiv
        \mathrm{QFT} \cdot
        U_{\mathrm{shift}} (d) \cdot
        \mathrm{QFT}^\dagger
    .
\end{align}
When this operator acts on the $n$ qubits responsible for the variable $x_0,$
it displaces the function $f$ by $d \Delta x$ in the $x_0$ direction as follows:
\begin{align}
    | f \rangle
    &\xmapsto{\mathrm{QFT}^\dagger}
        \sum_{j_0, \dots, j_{m - 1}}^{N - 1}
            f (j_0 \Delta x, j_1 \Delta x, \dots)
        \frac{1}{\sqrt{N}}
        \sum_{k_0 = 0}^{N - 1}
        \exp \frac{-2 \pi i j_0 k_0}{N}
        | k_0 \rangle_n
        \otimes
        | j_1 \rangle_n
        \otimes
        \cdots
        \otimes
        | j_{m - 1} \rangle_n
    \nonumber \\
    &\xmapsto{U_{\mathrm{shift}} (d)}
        \sum_{j_0, \dots, j_{m - 1}}^{N - 1}
            f (j_0 \Delta x, j_1 \Delta x, \dots)
        \frac{1}{\sqrt{N}}
        \sum_{k_0 = 0}^{N - 1}
        \exp \frac{2 \pi i (-j_0 - d) k_0}{N}
        | k_0 \rangle_n
        \otimes
        | j_1 \rangle_n
        \otimes
        \cdots
        \otimes
        | j_{m - 1} \rangle_n
    \nonumber \\
    &\xmapsto{\mathrm{QFT}}
        \sum_{j_0, \dots, j_{m - 1}}^{N - 1}
            f (j_0 \Delta x, j_1 \Delta x, \dots)
        \frac{1}{N}
        \sum_{k_0,k_0' = 0}^{N - 1}
        \exp \frac{2 \pi i (-j_0 - d + k_0') k_0}{N}
        | k_0' \rangle_n
        \otimes
        | j_1 \rangle_n
        \otimes
        \cdots
        \otimes
        | j_{m - 1} \rangle_n
    \nonumber \\
    &=
        \sum_{j_0, \dots, j_{m - 1}}^{N - 1}
            f (j_0 \Delta x, j_1 \Delta x, \dots)
        | j_0 + d \ \mathrm{mod} \ N \rangle_n
        \otimes
        | j_1 \rangle_n
        \otimes
        \cdots
        \otimes
        | j_{m - 1} \rangle_n
    \nonumber \\
    &=
        | f \ \mathrm{shifted \ by} \ d \Delta x \ \mathrm{in} \ x_0 \rangle
        .
\end{align}
This is also the case for the other variables.

For a vector $\boldsymbol{d}$ specified by $m$ integers,
we introduce the circuit $\mathcal{C}^{(\phi)} [\boldsymbol{d}]$
with an angle parameter $\phi,$
as shown in Fig.~\ref{circuit:circuit_for_derivative}.
The state of input $(mn + 1)$-qubit system consisting of the encoded function and the ancilla changes until the measurement as
\begin{align}
    | f \rangle
    \otimes
    \overbrace{
    | 0 \rangle
    }^{\mathrm{Ancilla}}
    \xmapsto{}
        \frac{
            | f \rangle
            +
            e^{i \phi}
            | f [\boldsymbol{d}] \rangle
        }{2}
        \otimes
        | 0 \rangle
        +
        \frac{
            | f \rangle
            -
            e^{i \phi}
            | f [\boldsymbol{d}] \rangle
        }{2}
        \otimes
        | 1 \rangle
    \equiv
        | \Psi [\boldsymbol{d}] \rangle
        ,
    \label{deriv_of_func_as_many_qubits:state_before_measurement}
\end{align}
where $| f [\boldsymbol{d}] \rangle$
represents $f$ displaced by
$\boldsymbol{h} \equiv \boldsymbol{d} \Delta x.$
The measurement is described by the 
$2 N^m$ projection operators
\begin{align}
    \mathcal{P}_{\boldsymbol{x} q}
    \equiv
        | \boldsymbol{x} \rangle
        \langle \boldsymbol{x} |
        \otimes
        | q \rangle \langle q |
    \
    (x_0, \dots, x_{m - 1} = 0, \Delta x, \dots, (N - 1) \Delta x, q = 0, 1)
    .
\end{align}
The probability for obtaining
$| \boldsymbol{x} \rangle \otimes | 0 \rangle$
as the outcome of measurement is thus,
from
Eq.~(\ref{deriv_of_func_as_many_qubits:state_before_measurement}),
\begin{align}
    \mathbb{P}_{\boldsymbol{x} 0} (\boldsymbol{d})
    &=
        \left\|
            \mathcal{P}_{\boldsymbol{x} 0}
            | \Psi [\boldsymbol{d}] \rangle
        \right\|^2
    \nonumber \\
    &=
        \frac{
            | f (\boldsymbol{x}) |^2
            +
            | f (\boldsymbol{x} - \boldsymbol{h}) |^2
        }{4}
        +
        \frac{\cos \phi}{2}
            \mathrm{Re}
            \left(
            f (\boldsymbol{x})
            f (\boldsymbol{x} - \boldsymbol{h})^*
            \right)
        +
        \frac{\sin \phi}{2}
            \mathrm{Im}
            \left(
            f (\boldsymbol{x})
            f (\boldsymbol{x} - \boldsymbol{h})^*
            \right)
\end{align}
This expression means that
we can obtain the relation which connects
the probability for $\phi = 0 \ (\phi = \pi/2)$
and the real (the imaginary) part of
$f (\boldsymbol{x}) f (\boldsymbol{x} - \boldsymbol{h})^*:$
\begin{align}
    f (\boldsymbol{x})
    f (\boldsymbol{x} - \boldsymbol{h})^*
    &=
        2
        \left(
            \mathbb{P}_{\boldsymbol{x} 0}^{(\mathrm{real})}
            (\boldsymbol{d})
            +
            i
            \mathbb{P}_{\boldsymbol{x} 0}^{(\mathrm{imag})}
            (\boldsymbol{d})
        \right)
        -
        \left(
            \mathbb{P}_{\boldsymbol{x}}
            +
            \mathbb{P}_{\boldsymbol{x} - \boldsymbol{h}}
        \right)
        \frac{1 + i}{2}
    .
\end{align}
The definition in 
Eq.~(\ref{deriv_of_func_as_many_qubits:def_G})
of the desired quantity
allows us to calculate it only from the probabilities as
\begin{gather}
    G^{(r)} (\boldsymbol{x}, \boldsymbol{h})
    =
        \begin{cases}
            \mathbb{P}_{\boldsymbol{x} 0}^{(\mathrm{real})}
            (-\boldsymbol{d})
            +
            i
            \mathbb{P}_{\boldsymbol{x} 0}^{(\mathrm{imag})}
            (-\boldsymbol{d})
            +
            \mathbb{P}_{\boldsymbol{x} 0}^{(\mathrm{real})}
            (\boldsymbol{d})
            +
            i
            \mathbb{P}_{\boldsymbol{x} 0}^{(\mathrm{imag})}
            (\boldsymbol{d})
            -
            \frac{1 + i}{4}
            (
                2
                \mathbb{P}_{\boldsymbol{x}}
                +
                \mathbb{P}_{\boldsymbol{x} + \boldsymbol{h}}
                +
                \mathbb{P}_{\boldsymbol{x} - \boldsymbol{h}}
            )
            &
            \mathrm{even} \ r
            \\
            \mathbb{P}_{\boldsymbol{x} 0}^{(\mathrm{real})}
            (-\boldsymbol{d})
            +
            i
            \mathbb{P}_{\boldsymbol{x} 0}^{(\mathrm{imag})}
            (-\boldsymbol{d})
            -
            \mathbb{P}_{\boldsymbol{x} 0}^{(\mathrm{real})}
            (\boldsymbol{d})
            -
            i
            \mathbb{P}_{\boldsymbol{x} 0}^{(\mathrm{imag})}
            (\boldsymbol{d})
            -
            \frac{1 + i}{4}
            (
                \mathbb{P}_{\boldsymbol{x} + \boldsymbol{h}}
                -
                \mathbb{P}_{\boldsymbol{x} - \boldsymbol{h}}
            )
            &
            \mathrm{odd} \ r
        \end{cases}
    .
    \label{deriv_of_func_as_many_qubits:G_from_measurements}
\end{gather}
In short,
we can obtain the derivative distributions
by solving the linear equation in
Eq.~(\ref{deriv_of_func_as_many_qubits:lin_eq_for_even_and_odd})
on a classical computer
by collecting data of $G^{(r)}$
for various $\boldsymbol{d}$ vectors
from repeated measurements on a quantum computer.
It should be stressed here that
this method does not involve the complex value of $f$ itself.
(In fact, we cannot know the shape of this complex function from this method.)
This characteristics of the method is favorable from a practical viewpoint since
the prepared $| f \rangle$ state for each measurement is not guaranteed to have a global phase common to all the measurements.

\subsection{An example for $m = 2$}

Let us consider a case where $f$ is a two-variable function $(m = 2).$
Eqs.~(\ref{deriv_of_func_as_many_qubits:num_of_integers_in_D})
and 
(\ref{deriv_of_func_as_many_qubits:num_of_unknowns_in_lin_eq})
give $N^{(1)} = 2$
for the first-order derivative distributions.
If we adopt, for example,
two displacement vectors
$\boldsymbol{h}_0 \equiv \boldsymbol{e}_0 \Delta x$
and
$\boldsymbol{h}_1 \equiv \boldsymbol{e}_1 \Delta x$
($\boldsymbol{e}_0$ and $\boldsymbol{e}_1$ are
the unit vector in the $x_0$ and $x_1$ directions,
respectively),
the linear equation is constructed from
Eq.~(\ref{deriv_of_func_as_many_qubits:lin_eq_for_even_and_odd})
as
\begin{align}
    \begin{pmatrix}
        G^{(1)} (\boldsymbol{x}, \boldsymbol{h}_0) \\
        G^{(1)} (\boldsymbol{x}, \boldsymbol{h}_1)
    \end{pmatrix}
    =
        \begin{pmatrix}
            \Delta x & 0 \\
            0 & \Delta x
        \end{pmatrix}
        \begin{pmatrix}
            g^{(1)}_{1, 0} (\boldsymbol{x}) \\
            g^{(1)}_{0, 1} (\boldsymbol{x}) \\
        \end{pmatrix}
        .
\end{align}
For the second-order distributions, on the other hand,
we have $N^{(2)} = 4.$
If we adopt, for example,
four displacement vectors
$
\boldsymbol{h}_0' \equiv \boldsymbol{e}_0 \Delta x,
\boldsymbol{h}_1' \equiv \boldsymbol{e}_1 \Delta x,
\boldsymbol{h}_2' \equiv (\boldsymbol{e}_0 + \boldsymbol{e}_1 ) \Delta x,
\boldsymbol{h}_2' \equiv (\boldsymbol{e}_0 - \boldsymbol{e}_1 ) \Delta x
,
$
the linear equation to be solved is constructed as
\begin{align}
    \begin{pmatrix}
        G^{(2)} (\boldsymbol{x}, \boldsymbol{h}_0') \\
        G^{(2)} (\boldsymbol{x}, \boldsymbol{h}_1') \\
        G^{(2)} (\boldsymbol{x}, \boldsymbol{h}_2') \\
        G^{(2)} (\boldsymbol{x}, \boldsymbol{h}_3')
    \end{pmatrix}
    =
        \begin{pmatrix}
            1 & \Delta x^2 & 0 & 0 \\
            1 & 0 & 0 & \Delta x^2 \\
            1 & \Delta x^2 & \Delta x^2 & \Delta x^2 \\
            1 & \Delta x^2 & -\Delta x^2 & \Delta x^2 \\
        \end{pmatrix}
        \begin{pmatrix}
            g^{(0)}_{0, 0} (\boldsymbol{x}) \\
            g^{(2)}_{2, 0} (\boldsymbol{x}) \\
            g^{(2)}_{1, 1} (\boldsymbol{x}) \\
            g^{(2)}_{0, 2} (\boldsymbol{x}) \\
        \end{pmatrix}
    .
\end{align}

\begin{figure}
\begin{center}
\includegraphics[width=9cm]{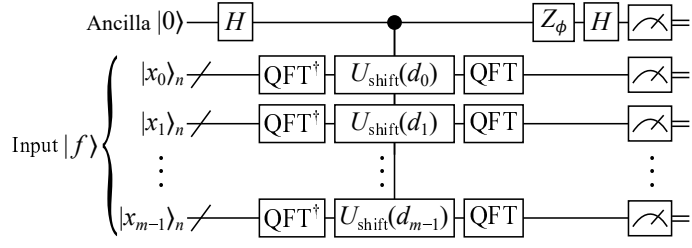}
\end{center}
\caption{
$(m n + 1)$-qubit circuit
$\mathcal{C}^{(\phi)} [\boldsymbol{d}]$
for obtaining the real or imaginary part of
$
f (\boldsymbol{x})
f (\boldsymbol{x} - \boldsymbol{h})^*
.
$
$
Z_\phi
\equiv
| 0 \rangle \langle 0 |
+
e^{i \phi} | 1 \rangle \langle 1 |
$
is a single-qubit phase gate.
$\mathcal{C}^{(\phi = 0)} [\boldsymbol{d}]$
is used for obtaining the real part,
while 
$\mathcal{C}^{(\phi = \pi/2)} [\boldsymbol{d}]$
is for the imaginary part.
}
\label{circuit:circuit_for_derivative}
\end{figure}

\section{Derivation of Eq.~(\ref{current_dens_paramag_x_from_meas})}
\label{sec:current_dens_paramag_x_from_meas}

We can assume that the index of the single electron on which
the circuit $\mathcal{C}_{\mathrm{para}, x} (d)$ 
in Fig.~\ref{fig:para_current_density_x}
acts is 0 without loss of generality.

The paramagnetic current density $j_{\mathrm{para}, x}$ can be seen as a
special case of the derivative distribution
already explained in Appendix
\ref{sec:fdfc_distr_from_measurements}
for $3 n_e$ variables and $r = 1.$
Specifically,
from Eq.~(\ref{current_dens_paramag_using_RDM1}),
the unknown distribution in this case is
$
\mathrm{Im} \partial \gamma
(\boldsymbol{r}, \boldsymbol{r}')/\partial x |_{\boldsymbol{r}' = \boldsymbol{r}}
=
-
\mathrm{Im} \partial \gamma
(\boldsymbol{r}', \boldsymbol{r})/\partial x |_{\boldsymbol{r}' = \boldsymbol{r}}.
$
That has the relation to the many-electron wave function $\Psi$ as
[see Eq.~(\ref{def_reduced_dens_mat_1e})]
\begin{align}
    \frac{\partial \gamma (\boldsymbol{r}', \boldsymbol{r})}
        {\partial x}
    \propto
        \int
        d^3 \boldsymbol{r}_1
        \cdots
        d^3 \boldsymbol{r}_{n_e - 1}
        \Psi (
            \boldsymbol{r}',
            \boldsymbol{r}_1,
            \dots,
            \boldsymbol{r}_{n_e - 1}
        )
        \frac{
        \partial
        \Psi (
            \boldsymbol{r},
            \boldsymbol{r}_1,
            \dots,
            \boldsymbol{r}_{n_e - 1}
        )^*
        }{\partial x}
        ,
\end{align}
meaning that $\Psi$ corresponds to $f$ in
Eq.~(\ref{deriv_of_func_as_many_qubits:def_g}).
The circuit $\mathcal{C}^{(\phi = \pi/2)} [\boldsymbol{d}]$
in Fig.~\ref{circuit:circuit_for_derivative}
with
a displacement vector $\boldsymbol{d} = \boldsymbol{e}_{0 x} d \Delta x$
($\boldsymbol{e}_{0 x}$ is a $3 n_e$-dimensional unit vector along the $x$ direction for the 0th electron) 
is almost equivalent to 
$\mathcal{C}_{\mathrm{para}, x} (d)$ in
Fig.~\ref{fig:para_current_density_x}.
These two circuits differ only in that
the $3 n (n_e - 1)$ qubits for electrons are not measured in $\mathcal{C}_{\mathrm{para}, x} (d),$
while all the $3 n n_e$ qubits are measured in
$\mathcal{C}^{(\phi = \pi/2)} [\boldsymbol{d}].$
This difference is correctly taken into account
for the RHS of
Eq.~(\ref{deriv_of_func_as_many_qubits:G_from_measurements})
by tracing out the degrees of freedom of the unobserved
$n_e - 1$ electrons as
\begin{gather}
    \sum_{\boldsymbol{x}_1, \dots, \boldsymbol{x}_{n_e - 1}}
    \left(
            \mathbb{P}_{\boldsymbol{x} 0}^{(\mathrm{imag})}
            (-\boldsymbol{d})
            -
            \mathbb{P}_{\boldsymbol{x} 0}^{(\mathrm{imag})}
            (\boldsymbol{d})
            -
            \frac{1}{4}
            (
                \mathbb{P}_{\boldsymbol{x} + \boldsymbol{h}}
                -
                \mathbb{P}_{\boldsymbol{x} - \boldsymbol{h}}
            )
    \right)
    \nonumber \\
    =
        \mathbb{P}_{\boldsymbol{x}_0 0}
        (-d)
        -
        \mathbb{P}_{\boldsymbol{x}_0 0}
        (d)
        -
        \frac{\Delta V}{4 n_e}
        \left(
            \rho
            (\boldsymbol{x}_0 + d \Delta x \boldsymbol{e}_x )
            -
            \rho
            (\boldsymbol{x}_0 - d \Delta x \boldsymbol{e}_x )
        \right)
    ,
\end{gather}
where $\mathbb{P}_{\boldsymbol{x}_0 0} (\pm d)$
are the probabilities in terms of
$\mathcal{C}_{\mathrm{para}, x} (\pm d).$
Since the quantity in the expression above
constitutes the linear equation for
$
\mathrm{Im} \partial \gamma
(\boldsymbol{r}', \boldsymbol{r})/\partial x |_{\boldsymbol{r}' = \boldsymbol{r}}
,
$
we get to Eq.~(\ref{current_dens_paramag_x_from_meas})
for $j_{\mathrm{para}, x}.$

\section{Derivation of Eqs.~(\ref{eigenstate_canceller:success_state_for_filt1}) and (\ref{eigenstate_canceller:success_state_for_filt2})}
\label{sec:derivation_for_filtration}

\subsection{For first-order filtration}

We expand the input state in terms of the energy eigenstates as
$| \psi \rangle = \sum_k c_k | \phi_k \rangle$
for the filtration circuit 
$\mathcal{C}_{\mathrm{filt1}}^{(\widetilde{E}_0)}.$
Substitution of this into the $(n + 1)$-qubit state in 
Eq.~(\ref{eigenstate_canceller:action_of_filt})
for an arbitrary real-time step $\Delta t_{\mathrm{f}}$ leads to
\begin{gather}
    \sum_k
        c_k
        \left(
            \frac{
            e^{-i \widetilde{E}_0 \Delta t_{\mathrm{f}} /2}
            -
            e^{i (\widetilde{E}_0/2 - E_k) \Delta t_{\mathrm{f}} }}{2}
            | \phi_k \rangle
            \otimes
            | 0 \rangle
            +
            \frac{
            e^{-i \widetilde{E}_0 \Delta t_{\mathrm{f}} /2}
            +
            e^{i (\widetilde{E}_0/2 - E_k) \Delta t_{\mathrm{f}} }}{2}
            | \phi_k \rangle
            \otimes
            | 1 \rangle       
        \right)
    \nonumber \\
    =
    \sum_k
        c_k
        e^{-i E_k \Delta t_{\mathrm{f}}/2}
        \left(
            i
            \sin \frac{(E_k - \widetilde{E}_0) \Delta t_{\mathrm{f}}}{2}
            | \phi_k \rangle
            \otimes
            | 0 \rangle
            +
            \cos \frac{(E_k - \widetilde{E}_0) \Delta t_{\mathrm{f}}}{2}
            | \phi_k \rangle
            \otimes
            | 1 \rangle       
        \right)
        .
\end{gather}
As we set $\Delta t_{\mathrm{f}}$ to $\Delta t_1,$
defined in Eq.~(\ref{eigenstate_canceller:optimal_dt}),
the unnormalized success state in the equation above is written as
\begin{gather}
    c_0
    \exp
    \left(
        -i
        \frac{\pi E_0}
        {2 (\widetilde{E}_1 - \widetilde{E}_0)}
    \right)
    \sin
    \frac{\pi \delta E_0}
    {2 (\widetilde{E}_1 - \widetilde{E}_0)}
    | \phi_0 \rangle
    +
    \sum_{k \ne 0}
        c_k
        \exp
        \left(
            -i
            \frac{\pi E_k}
            {2 (\widetilde{E}_1 - \widetilde{E}_0)}
        \right)
        \sin
        \frac{\pi (E_k - \widetilde{E}_0)}
        {2 (\widetilde{E}_1 - \widetilde{E}_0)}
        | \phi_k \rangle
    ,
\end{gather}
where the amplitude of ground state is clearly $\mathcal{O} (\delta E_0).$
If the estimated energy eigenvalues of the ground and the first-excited states have no errors $(\delta E_0 = \delta E_1 = 0),$
the weight of $| \phi_1 \rangle$ in the equation above is
$| c_k |^2 \sin^2 (\pi/2) = | c_k |^2,$
achieving the maximum of $\sin^2$ factor,
which explains the appropriateness of $\Delta t_1.$

\subsection{For second-order filtration}

We substitute the input state $| \psi \rangle$
expanded in terms of the energy eigenstate
for the filtration circuit
$\mathcal{C}_{\mathrm{filt2}}^{(\widetilde{E}_0)}$
into
Eq.~(\ref{eigenstate_canceller:action_of_filt_2})
for an arbitrary $\Delta t_{\mathrm{f}}$ to get the unnormalized success state
\begin{gather}
    \sum_k
        c_k
        \sin^2
        \frac{(E_k - \widetilde{E}_0) \Delta t_{\mathrm{f}}}{2}
        | \phi_k \rangle
        .
\end{gather}
As we set $\Delta t_{\mathrm{f}}$ to $\Delta t_1,$
this state is written as
\begin{gather}
    c_0
    \sin^2
    \frac{\pi \delta E_0 }{2 (\widetilde{E}_1 - \widetilde{E}_0)}
    | \phi_0 \rangle
    +
    \sum_{k \ne 0}
        c_k
        \sin^2
        \frac{\pi (E_k - \widetilde{E}_0) }{2 (\widetilde{E}_1 - \widetilde{E}_0)}
        | \phi_k \rangle
        ,
\end{gather}
where the amplitude of ground state is clearly $\mathcal{O} (\delta E_0^2).$
If the estimated energy eigenvalues of the ground and the first-excited states have no errors,
the weight of $| \phi_1 \rangle$ in the equation above
achieves the maximum as well as in the first-order filtration.

\section{Matrix elements of a single-particle Hamiltonian}
\label{sec:Hamiltonian_mat_of_1p}

Here we provide the expression of matrix elements of a generic single-particle Hamiltonian.

The matrix element of the kinetic-energy operator in
Eq.~(\ref{def_kinetic_opr}),
or equivalently the kinetic propagator,
between the position eigenstates
$| \boldsymbol{r}^{(\boldsymbol{k})} \rangle$ and
$| \boldsymbol{r}^{(\boldsymbol{k}')} \rangle$
is calculated as
\begin{align}
    \langle \boldsymbol{r}^{(\boldsymbol{k})} |
    \hat{T}
    | \boldsymbol{r}^{(\boldsymbol{k}')} \rangle
    &=
        \frac{1}{2 m}
        \langle \boldsymbol{r}^{(\boldsymbol{k})} |
            \left(
                \hat{\boldsymbol{p}}
                -
                \frac{q}{c}
                \boldsymbol{A} (\boldsymbol{r}^{(\boldsymbol{k})})
            \right)
            \cdot
            \left(
                \hat{\boldsymbol{p}}
                -
                \frac{q}{c}
                \boldsymbol{A} (\boldsymbol{r}^{(\boldsymbol{k}')})
            \right)
        | \boldsymbol{r}^{(\boldsymbol{k}')} \rangle
    \nonumber \\
    &=
        \frac{1}{2 m}
        \langle \boldsymbol{r}^{(\boldsymbol{k})} |
            \left(
                \hat{\boldsymbol{p}}
                -
                \frac{q}{c}
                \boldsymbol{A} (\boldsymbol{r}^{(\boldsymbol{k})})
            \right)
            \cdot
            \left(
                \hat{\boldsymbol{p}}
                -
                \frac{q}{c}
                \boldsymbol{A} (\boldsymbol{r}^{(\boldsymbol{k}')})
            \right)
            \left(
                \sum_{s_x, s_y, s_z = 0}^{N - 1}
                | \boldsymbol{p}^{(\widetilde{\boldsymbol{s}})} \rangle
                \langle \boldsymbol{p}^{(\widetilde{\boldsymbol{s}})} |
            \right)
        | \boldsymbol{r}^{(\boldsymbol{k}')} \rangle
    \nonumber \\
    &=
        \frac{1}{N^3}
        \sum_{s_x, s_y, s_z = 0}^{N - 1}
            \frac{1}{2 m}
            \left(
                \boldsymbol{p}^{(\widetilde{\boldsymbol{s}})}
                -
                \frac{q}{c}
                \boldsymbol{A} (\boldsymbol{r}^{(\boldsymbol{k})})
            \right)
            \cdot
            \left(
                \boldsymbol{p}^{(\widetilde{\boldsymbol{s}})}
                -
                \frac{q}{c}
                \boldsymbol{A} (\boldsymbol{r}^{(\boldsymbol{k}')})
            \right)
        \exp
        \left(
            i
            \boldsymbol{p}^{(\widetilde{\boldsymbol{s}})}
            \cdot
            \left(
                \boldsymbol{r}^{(\boldsymbol{k})} 
                -
                \boldsymbol{r}^{(\boldsymbol{k}')} 
            \right)
        \right)
    \nonumber \\
    &=
        \frac{1}{N^3}
        \sum_{s_x, s_y, s_z = 0}^{N - 1}
        \sum_{\nu = x, y, z}
            \frac{1}{2 m}
            \Pi_{\nu \boldsymbol{k} \boldsymbol{s}}
            \Pi_{\nu \boldsymbol{k}' \boldsymbol{s}}^*
    ,
\end{align}
where the second equality was obtained by employing
the fact that the momentum eigenstates in
Eq.~(\ref{QITE_as_a_part_of_RITE:def_mom_eigenstate})
form a complete orthonormal system for $3 n$-qubit states.
$\Pi_\nu$ is a $N^3$-dimensional non-hermitian matrix.
Its $(\boldsymbol{k}, \boldsymbol{s})$ component is defined as
\begin{gather}
    \Pi_{\nu \boldsymbol{k} \boldsymbol{s}}
    \equiv
        \left(
            p_\nu^{(\widetilde{\boldsymbol{s}})}
            -
            \frac{q}{c}
            A_\nu (\boldsymbol{r}^{(\boldsymbol{k})})
        \right)
        \exp
        \left(
            i
            \boldsymbol{p}^{(\widetilde{\boldsymbol{s}})}
            \cdot
            \boldsymbol{r}^{(\boldsymbol{k})} 
        \right)
        .
\end{gather}

The matrix element of the potential operator is given by
\begin{gather}
    \langle \boldsymbol{r}^{(\boldsymbol{k})} |
    \hat{V}
    | \boldsymbol{r}^{(\boldsymbol{k}')} \rangle
    =
        V (\boldsymbol{r}^{(\boldsymbol{k})})
        \delta_{\boldsymbol{k} \boldsymbol{k}'}
        .
\end{gather}

\section{Energy spectra}

\subsection{For the harmonic potential in Sect.~\ref{sec:results:harmonic}}
\label{sec:spectra_for_harmonic}

The lowest ten energy eigenvalues of the Fock--Darwin system
obtained by numerical diagonalization for
the simulation cell size $L = 120$ nm for
12 qubits (six for each of $x$ and $y$ directions) are shown as circles in 
Fig.~\ref{fig:single_well_spectra}
as functions of the external magnetic field $B.$
By using the cyclotron frequency
$\omega_{\mathrm{c}} \equiv \sqrt{|e B|/(m c)}$
of the electron, having the charge $e$,
and $\Omega \equiv \sqrt{\omega_0^2 + \omega_{\mathrm{c}}^2/4}$,
the analytic expression for the energy eigenvalues is known to be given by
two quantum numbers as
$E_{n_1 \ell} = (n_1 + 1) \Omega - \ell \omega_{\mathrm{c}}/2$
for $n_1 = 0, 1, \dots$ and
\begin{align}
    \ell
    =
    \begin{cases}
        0, \pm 2, \pm 4, \dots, \pm n_1 & \mathrm{for \ even} \ n_1
        \\
        \pm 1, \pm 3, \dots, \pm n_1 & \mathrm{for \ odd} \ n_1
    \end{cases}
    .
    \label{analytic_energy_eval_of_Fock_Darwin}
\end{align}
The analytically obtained eigenvalues for $n_1 \leq 3$ are also shown as the solid curves in the figure.

\begin{figure}
\begin{center}
\includegraphics[width=6.5cm]{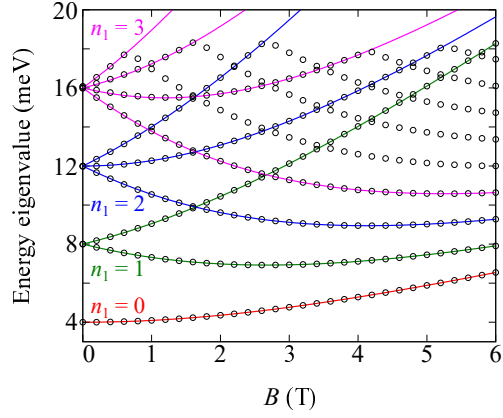}
\end{center}
\caption{
Circles represent the lowest ten energy eigenvalues
of the Fock--Darwin system for each of $B$
obtained by numerical diagonalization of the Hamiltonian.
Solid curves represent those obtained by the analytic expression
in Eq.~(\ref{analytic_energy_eval_of_Fock_Darwin}),
specified by quantum numbers $n_1$ and $\ell.$
}
\label{fig:single_well_spectra}
\end{figure}

\subsection{For the double-well potential in Sect.~\ref{sec:results:double_well}}
\label{sec:spectra_for_double_well}

The lowest energy eigenvalues of an electron for the double-well potential in Eq.~(\ref{def_double_well_pot})
obtained by numerical diagonalization 
are shown as circles in Fig.~\ref{fig:double_well_spectra}
as functions of the external magnetic field $B.$
The simulation cell size $L = 120$ nm for
12 qubits (six for each of $x$ and $y$ directions) 
was used.

\begin{figure}
\begin{center}
\includegraphics[width=6.5cm]{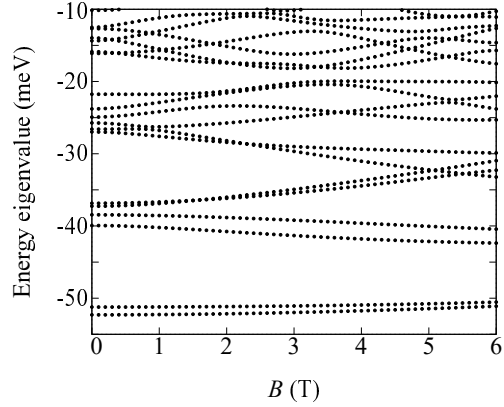}
\end{center}
\caption{
Circles represent the single-electron energy spectra
for the double-well potential for each of $B$
obtained by numerical diagonalization of the Hamiltonian.
}
\label{fig:double_well_spectra}
\end{figure}

\section{CNOT gate counts for the PITE circuits in the simulations}
\label{sec:gate_counts}

Here we consider the CNOT gate counts for the PITE circuits used in our simulations.
They are good measures of the computational cost
since it is well known that two-qubit gates are in general much prone to errors and spend much longer time than single-qubit gates on a real quantum computer.

\subsection{Expressions for generic potentials}

Since the simulations are for a single electron in two-dimensional space,
the number of CNOT gates in a single PITE step
using the $TV$ splitting is calculated as
\begin{align}
    c^{(TV)}
    =
        c (U_{\mathrm{pot}})
        +
        c (\mathrm{C} U_{\mathrm{pot}})
        +
        6
        c (\mathrm{QFT})
        +
        2
        c (U_{\mathrm{kin}})
        +
        2
        c (\mathrm{C} U_{\mathrm{kin}})
        +
        2
        c (U_{\mathrm{mag}})
        ,
    \label{CNOTs_for_TV}
\end{align}
where we used the fact that each phase gate and its squared form have the same circuit structure (see also Fig.~\ref{fig:circuit_for_electrons}).
From Figs.~\ref{fig:generic_approx_circuit}(c) and
\ref{fig:circuits_for_kinetic}(a),
that for the $TVT$ splitting is similarly calculated as
\begin{align}
    c^{(TVT)}
    =
        c (U_{\mathrm{pot}})
        +
        c (\mathrm{C} U_{\mathrm{pot}})
        +
        14
        c (\mathrm{QFT})
        +
        4
        c (U_{\mathrm{kin}})
        +
        4
        c (\mathrm{C} U_{\mathrm{kin}})
        +
        6
        c (U_{\mathrm{mag}})
        .
    \label{CNOTs_for_TVT}
\end{align}
Recalling that any singly-controlled single-qubit operation can be implemented by using two CNOT gates \cite{bib:5029, Nielsen_and_Chuang},
we find that the ordinary implementation of QFT \cite{Nielsen_and_Chuang} involves 
$c (\mathrm{QFT}) = n^2 + n/2$ CNOT gates for $n = 6$ in our simulations.
From Eq.~(\ref{PITE_with_mag_fields:impl_of_U_mag}),
the gate count for the magnetic-phase gate is 
$c (U_{\mathrm{mag}}) = 2 n^2.$

From Eq.~(\ref{PITE_with_mag_fields:impl_of_U_kin}),
$U_{\mathrm{kin}}$ contains $n (n - 1)/2$ controlled phase gates.
Its gate count is thus
$c (U_{\mathrm{kin}}) = n (n - 1).$
The controlled kinetic-phase gate consists of
$n$ singly controlled single-qubit gates and
$c (U_{\mathrm{kin}})$ doubly controlled single-qubit gates.
It is known that any doubly controlled single-qubit gate can be implemented by using six CNOT gates \cite{bib:5029, Nielsen_and_Chuang}.
Thus we have 
$c (\mathrm{C} U_{\mathrm{kin}}) = 3 n^2 - n.$

Gathering the observations above,
we can rewrite
Eqs.~(\ref{CNOTs_for_TV}) and (\ref{CNOTs_for_TVT}) as
\begin{align}
    c^{(TV)}
    =
        c (U_{\mathrm{pot}})
        +
        c (\mathrm{C} U_{\mathrm{pot}})
        +
        18 n^2 - 2 n
\label{CNOTs_for_TV_using_n}
\end{align}
and
\begin{align}
    c^{(TVT)}
    =
        c (U_{\mathrm{pot}})
        +
        c (\mathrm{C} U_{\mathrm{pot}})
        +
        42 n^2 - n
        ,
\label{CNOTs_for_TVT_using_n}
\end{align}
respectively.

\subsection{For the harmonic potential in Sect.~\ref{sec:results:harmonic}}

The potential-evolution operator in this case is factorized as
$
U_{\mathrm{pot}} (\Delta t)
=
\exp
(-i m \omega_0^2 (x - L/2)^2 \Delta t/2)
\exp
(-i m \omega_0^2 (y - L/2)^2 \Delta t/2)
.
$
Each of the two operators on the right-hand side has a similar functional form to $U_{\mathrm{kin}} (\Delta t)$
[see Eq.~(\ref{PITE_with_mag_fields:action_of_U_kin})].
Thus we have the gate counts
$c (U_{\mathrm{pot}}) = 2 c (U_{\mathrm{kin}})$
and
$
c (\mathrm{C} U_{\mathrm{pot}})
=
2 c (\mathrm{C} U_{\mathrm{kin}})
.
$
From Eqs.~(\ref{CNOTs_for_TV_using_n}) and
(\ref{CNOTs_for_TVT_using_n}),
the gate counts for the entire PITE step are calculated as
$c^{(TV)} = 26 n^2 - 5 n = 906$ and
$c^{(TVT)} = 50 n^2 - 5 n = 1770.$

\subsection{For the double-well potential in Sect.~\ref{sec:results:double_well}}

There can be various implementation of the potential-evolution operator in this case.
We adopt here one possible implementation for which we consider the gate counts.
We introduce an $n_{\mathrm{a}}$-qubit auxiliary register to define the squared-distance unitary via
\begin{align}
    S (x_{\mathrm{c}}, d_x)
    \left(
        | x^{(k)} \rangle
        \otimes
        | 0 \rangle_{n_{\mathrm{a}}}
    \right)
    =
        | x^{(k)} \rangle
        \otimes
        \left|
            \frac{(x^{(k)} - x_{\mathrm{c}})^2}{d_x^2}
        \right\rangle_{n_{\mathrm{a}}}
    \label{impl_evol_for_double_well:def_sq_dist}
\end{align}
for each $k$ and $k',$
where the center $x_{\mathrm{c}}$ and the width $d_x$ are predetermined circuit parameters.
The state of the auxiliary register on the right-hand side of
Eq.~(\ref{impl_evol_for_double_well:def_sq_dist})
is one of the states $| j \rangle_{n_{\mathrm{a}}} \ (j = 0, \dots, 2^{n_{\mathrm{a}}} - 1)$ where $j$ corresponds to the discretized possible values of
$(x - x_{\mathrm{c}})^2/d_x^2.$
We assume that $n_{\mathrm{a}}$ is sufficiently large for accurate discretization of such continuous values.
We define the unitary adder via
\begin{align}
    \mathrm{ADD}
    \left(
        | r \rangle_{n_{\mathrm{a}}}
        \otimes
        | r' \rangle_{n_{\mathrm{a}}}
        \otimes
        | 0 \rangle_{n_{\mathrm{a}}}
    \right)
    =
        | r \rangle_{n_{\mathrm{a}}}
        \otimes
        | r' \rangle_{n_{\mathrm{a}}}
        \otimes
        | r + r' \rangle_{n_{\mathrm{a}}}
    .
    \label{impl_evol_for_double_well:def_adder}
\end{align}
Explicit construction of an adder is known \cite{bib:5379, bib:5396, bib:5397}.
We define the exponential-function phase gate via
\begin{align}
    U_e (\theta)
    | r \rangle_{n_{\mathrm{a}}}
    \equiv
        \exp (-i e^{-r} \theta)
        | r \rangle_{n_{\mathrm{a}}}
    \label{impl_evol_for_double_well:def_exp_func_phase}
\end{align}
with a predetermined circuit parameter $\theta.$
Although the unitary gates defined in
Eqs.~(\ref{impl_evol_for_double_well:def_sq_dist}),
(\ref{impl_evol_for_double_well:def_adder}), and
(\ref{impl_evol_for_double_well:def_exp_func_phase})
may involve ancillary qubits other than the auxiliary registers,
we do not write them explicitly in the equations for simplicity.

We define the Gaussian evolution circuit
$\mathcal{C}_{\mathrm{G}} (x_{\mathrm{c}}, y_{\mathrm{c}}, d_x, d_y, \theta)$
as shown in Fig.~\ref{fig:circuit_for_double_well},
which is made up of the three major parts:
the computation of the argument,
the evolution due to the potential,
the uncomputation.
One can easily confirm that this circuit implements the evolution for a Gaussian potential:
\begin{gather}
    \mathcal{C}_{\mathrm{G}}
    (x_{\mathrm{c}}, y_{\mathrm{c}}, d_x, d_y, \theta)
    \left(
        | x^{(k)} \rangle 
        \otimes
        | y^{(k')} \rangle 
        \otimes
        ( \mathrm{initialized \ auxiliary \ qubits} )
    \right)
    \nonumber \\
    =
        \exp
        \left(
            -i \theta
            \exp
            \left(
                -
                \frac{(x^{(k)} - x_{\mathrm{c}})^2}{d_x^2}
                -
                \frac{(y^{(k')} - y_{\mathrm{c}})^2}{d_y^2}
            \right)
        \right)
        | x^{(k)} \rangle 
        \otimes
        | y^{(k')} \rangle 
        \otimes
        ( \mathrm{initialized \ auxiliary \ qubits} )
        .
\end{gather}
Thanks to the uncomputation part,
which disentangles the auxiliary registers from the electronic register, 
we can discard the auxiliary registers safely \cite{Nielsen_and_Chuang}.
Since the potential in the simulation is the sum of three Gaussian functions [see Eq.~(\ref{def_double_well_pot})],
we can implement the potential-evolution as the consecutive Gaussian evolution circuits as
\begin{align}
    U_{\mathrm{pot}} (\Delta t)
    =
        \mathcal{C}_{\mathrm{G}}
        \left(
            \frac{L}{2}, \frac{L}{2},
            \Delta_x, \Delta_y, V_p \Delta t
        \right)
        \mathcal{C}_{\mathrm{G}}
        \left(
            \frac{L}{2} + a, \frac{L}{2},
            \Delta, \Delta, V_0 \Delta t
        \right)
        \mathcal{C}_{\mathrm{G}}
        \left(
            \frac{L}{2} - a,
            \frac{L}{2},
            \Delta, \Delta, V_0 \Delta t
        \right)
        .
\end{align}
The gate count for this implementation is thus given by
\begin{align}
    c (U_{\mathrm{pot}})
    &=
        3
        c (\mathcal{C}_{\mathrm{G}})
    \nonumber \\
    &=
        12
        c (S)
        +
        6
        c (\mathrm{ADD})
        +
        3
        c (U_e)
    .
    \label{impl_evol_for_double_well:gate_count_for_double_well_pot}
\end{align}
That for the controlled $U_{\mathrm{pot}}$ can also be calculated similarly.
Explicit construction of the circuit for $S (x_c, d_x)$
is possible since the arithmetic circuits for addition \cite{bib:5379, bib:5396, bib:5397} and multiplication \cite{bib:5405, bib:5398, bib:5399} are known.
That of the circuit for $U_e (\theta)$ is also possible by approximating the exponential function as a piecewisely defined polynomial \cite{bib:5389, bib:5384}.
We can thus evaluate
Eq.~(\ref{impl_evol_for_double_well:gate_count_for_double_well_pot}).
The explicit construction of those circuits, however,
goes beyond the scope of the present study.

\begin{figure}
\begin{center}
\includegraphics[width=8cm]{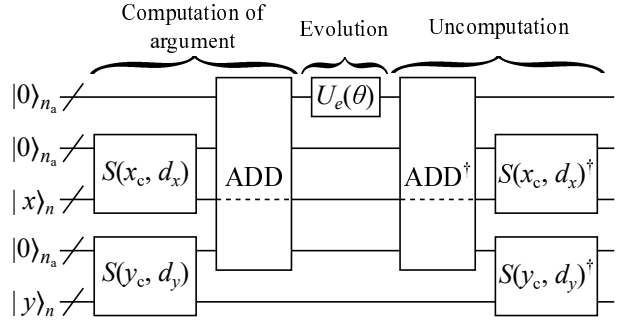}
\end{center}
\caption{
Gaussian evolution circuit
$
\mathcal{C}_{\mathrm{G}}
(x_{\mathrm{c}}, y_{\mathrm{c}}, d_x, d_y, \theta)
$
as a building block of $U_{\mathrm{pot}} (\Delta t)$ for
the double-well potential in Sect.~\ref{sec:results:double_well}.
}
\label{fig:circuit_for_double_well}
\end{figure}

\end{widetext}

\bibliography{ref}

\end{document}